\documentclass [12pt,twoside]{article}           
\countdef\arxiv=201

\ifnum\arxiv=0  \input cpreb.tex  \fi

\ifnum\arxiv=1

\usepackage{epsfig,times,lscape}
\usepackage[usenames]{color}

    \ifnum\count102=1

    \fi

    \ifnum\count102=2
\fi

\newcommand{\nc}{\newcommand}

     \ifnum\count101=1
\nc{\qI}[1]{\section{{#1}}}
\nc{\qA}[1]{\subsection{{#1}}}
\nc{\qun}[1]{\subsubsection{{#1}}}
\nc{\qa}[1]{\paragraph{{#1}}}

\def\qpar{\vskip 2mm plus 0.2mm minus 0.2mm}
\def\qL{\hfill \break}
     \fi 

      \ifnum\count101=2
 \nc{\qI}[1]{\parindent=0mm \vskip 8mm 
{\centerline{\LARGE \color{red}#1}}\vskip 3mm}
\nc{\qA}[1]{\vskip 2.5mm \noindent 
{{\bf\large\color{blue}  #1}} \vskip 1mm \parindent=0mm}
 \nc{\qun}[1]{\vskip 1mm \noindent {\sl #1 }\quad }

\def\qL{\hfill \break}
\def\qpar{\vskip 2mm plus 0.2mm minus 0.2mm}

      \fi

\def\qth{\vrule height 12pt depth 0pt width 0pt}
\def\qtb{\vrule height 0pt depth 5pt width 0pt}

\nc{\qfoot}[1]{\footnote{{#1}}}

      \ifnum\count101=1
\def\qbu{\hfill \par \hskip 6mm $ \bullet $ \hskip 2mm}

\def\qee#1{\hfill \par \hskip 6mm (#1) \hskip 2 mm}
\def\qeeb#1{\hfill \par \hskip 6mm {\color{blue} (#1)} \hskip 2 mm}
      \fi
      \ifnum\count101=2
\def\qbu{\hfill \par \hskip 4mm $ \bullet $ \hskip 2mm}

\def\qee#1{\hfill \par \hskip 4mm (#1) \hskip 2 mm}
\def\qeeb#1{\hfill \par \hskip 6mm {\color{blue} (#1)} \hskip 2 mm}
      \fi

\def\qparr{ \vskip 1.0mm plus 0.2mm minus 0.2mm \hangindent=10mm
\hangafter=1}

     \ifnum\count101=1 
 
     \fi
     \ifnum\count101=2

  \def\qcitb#1{\noindent \hbox to 102mm{\hfill \small #1} \vskip 1mm}
      \fi

 \def\qpages#1{\count102=0{\loop\advance\count102 by 1
 \null \vfill\eject \ifnum\count102<#1 \repeat}}

\def\qth{\vrule height 12pt depth 0pt width 0pt}
\def\qtb{\vrule height 0pt depth 5pt width 0pt}

\def\qv{\vskip 0.1mm plus 0.05mm minus 0.05mm}
\def\qhu{\hskip 0.6mm}
\def\qhv{\hskip 3mm}

\def\qhw{\hskip 1.5mm}
\def\qleg#1#2#3{\noindent {\bf \small #1\qhw}{\small #2\qhw}{\it \small #3}\qv }
\fi

\begin{document}
\thispagestyle{empty}



\markboth{{\sl \hfill  \hfill \protect\phantom{3}}}
        {{\protect\phantom{3}\sl \hfill  \hfill}}

\color{yellow} 
\hrule height 20mm depth 15mm width 170mm 
\color{black}
\vskip -2.7cm 
 \centerline{\bf \LARGE How can one measure group cohesion?}
\vskip 2mm
 \centerline{\bf \large From individual organisms to their interaction}
\vskip 4mm
\centerline{Z. Di$ ^1 $, M. Gho$ ^2 $, X. Lu$ ^3 $, G. Li$ ^4 $,
B.M. Roehner$ ^5 $, N.J. Suematsu$ ^6 $, C. Y\'epr\'emian$ ^7 $
}
\vskip 7mm
\hfil {\it Invited talk given at the COST}%
\qfoot{``COST'' means {\bf Co}operation in {\bf S}cience and 
{\bf T}echnology.
It is a program supported and funded by the European Union.
Entitled ``Ways of seeing'', the workshop lasted from 23 to
25 April 2014}
{\it Workshop held in Galway (Ireland), 25 April 2014, 9:30-10:30.}
\qfoot{This paper is also available on the ``arXiv'' (Princeton) 
and ``BOLA'' (Beijing) preprint archives.}%
. \hfil

\vskip 6mm
\normalsize
{\bf Abstract}\quad Measuring atomic and molecular interactions
was one of the main objectives of physics during the past century.
It was an essential step not only in itself but because
most macroscopic properties can be derived once one knows
interaction strengths. At the present time, except for
systems that can be described as discrete networks (like
the Internet network) 
our knowledge
of social and biological ties still remains very limited.
An important step is to develop experimental
means for measuring social and biological interactions.
In this talk there are two parts. 
\qbu Firstly, we describe
experimental evidence of inter-individual
attraction in populations of insects. 
\qbu Secondly, we 
focus on a specific system, namely populations of {\it Euglena
gracilis}, a green, swimming unicellular organism,
for which we try to determine individual and interaction 
properties.

\vskip 3mm
\centerline{\it Version of 13 June 2014. Comments are welcome.}

\vskip 5mm
{\bf \color{blue} Websites which may be useful in connection with 
this topic:} 
\qbu {http://www.lpthe.jussieu.fr/$ \sim $roehner/bola.html 
\quad
[Papers on the physics of living populations] 
\qbu http://www.lpthe.jussieu.fr/$ \sim $roehner/expclust.html \quad
[Pictures of clustering processes.]

\vskip 6mm

{\small 
1: Xengru Di, School of Systems Science, Beijing Normal University.\qL
\phantom{1: }Email: zdi@bnu.edu.cn \qL
2: Michel Gho, ``Cell cycle and cell determination'' group at University 
Pierre and Marie Curie, Paris.\qL
\phantom{1: }Email: Michel.Gho@snv.jussieu.fr\qL
3: Xia Lu, Genetics group of Pr. Dou Fei at Beijing Normal
University. \qL
\phantom{1: }Email: melody4ever9t@gmail.com\qL
4: Geng Li, School of Systems Science, Beijing Normal University.\qL
\phantom{1: }Email: moralmarket@126.com \qL
5: Bertrand Roehner, Institute for Theoretical and High Energy Physics 
(LPTHE), University Pierre and Marie Curie, Paris. \qL
\phantom{1: }Email: roehner@lpthe.jussieu.fr\qL
6: Nobuhiko Suematsu, Department of ``Mathematics and Applications'' of
Meiji University, Tokyo.\qL
\phantom{1: }Email: suematsu@meiji.ac.jp \qL
7: Claude Y\'epr\'emian, ``Mol\'ecules de communication et
adaptation des microorganismes'', Unit\'e Mixte de Recherche 7245 of
the ``Centre National pour la Recherche Scientifique'', Mus\'eum
National d'Histoire Naturelle, Paris.\qL
\phantom{1: }Email: yep@mnhn.fr \qL
}

\large

\vfill\eject


{\Large \bf Contents}
\vskip 3mm
{\bf Part 1: Introduction}
\qpar
Rationale for experiments on insects and microorganisms \qL
Characteristics of interactions: strength, range, duration \qL
Clustering as a means for estimating inter-attraction 
\qpar
{\bf Part 2: Examples of inter-attractive behavior}
\qpar
Clustering as a means for estimating inter-attraction\qL
Attraction strength among social insects: ants, bees \qL
Attraction strength among non-social insects: beetles, drosophila
\qpar
{\bf Part 3: Lessons and hints from physical chemistry}
\qpar
How to use aggregation/diffusion to measure interaction? \qL
Gravity, noise and interaction in diffusion experiments \qL
Application to the case of living organisms
\qpar
{\bf Part 4: In search of a general measurement method}
\qpar
{\it Euglena gracilis}
Research plan \qL
Individual properties \qL
Relationship between local density and speed \qL
Reactions brought about by light \qL
Evidence of interaction \qL
Network formation
\vskip 7mm

\centerline{\LARGE \bf  \color{magenta} Part 1: Introduction}
\vskip 3mm

\qI{Rationale for experiments on insects and microorganisms}

\qA{Group cohesion in physics}

What holds a system of particles 
together? What are the laws which rule
the transition from a state of high cohesion to a state of
lower cohesion and vice versa? 
\qpar

Such questions
have been of fundamental importance in physics
and chemistry. As an illustration, one can mention
that when in 1900 at the age of 21  
Albert Einstein submitted his first 
paper to the prestigious German journal ``Annalen der Physik''
his purpose was to derive the characteristics of 
inter-molecular attraction potentials from data about
capillarity properties of various chemicals. 
At that time,
such an objective was shared by many physicists.
\qpar

Then, during the 20th century,  little by little
our understanding and
knowledge of microscopic interactions 
progressively became more accurate. 
\qpar
Incidentally, it can be observed that 
in a sense, the fact that until the second half of the 20th century
molecular mechanisms were out of direct 
observational reach was a chance because it allowed physicists
and chemists to focus on global properties before 
becoming involved in the complicated details of
molecular mechanisms. 

\qA{Group cohesion in human systems}
The same questions constitute also central issues
for social systems. Yet, they are not 
often formulated in such a way. Consider the following examples.
\qbu It is commonly recognized that high performance teams or
companies are characterized by effective interactions, but how can 
such interactions be measured? One might be tempted to use
surveys; however such surveys will be unreliable when 
team members do not feel free to tell the truth, a situation
which is likely to occur when there are no trustful 
relations.  
\qbu  Everybody understands that the relationship
between parents and their children is (in some way
that needs to be defined)
``stronger'' than the connection they may have with 
neighbors or colleagues. The later may in turn be stronger
than the links they have with fellow citizens. 
Yet, we are unable to quantify
the respective strengths of such interactions. Why is this so?
Here is a possible answer. 
\qpar
Before one can talk about
quantification is it not necessary to define more
precisely what one wants to measure? In science
definition and measurement usually go hand in hand.
If we could set up an experimental procedure for
measuring the strength of family bonds the very conditions
of the experiment would spell out what the
experiment is measuring.
But can we set up an experiment involving parents, children
or citizens? Clearly not for obvious ethical reasons.
\qpar
However, it is possible to perform experiments on 
living organisms such as insects 
and microorganisms, all the more so because such
experiments do no require to kill or harm these organisms.
The present research is in line with previous 
pioneering studies based on
insect models such as Costa (1997),
Viswanathan et al. (2011),
Suematsu  et al. (2011), Nishimori (2012).

\qI{Characteristics of interactions: strength, range, duration}

Before trying to characterize interactions let us show a picture
which illustrates one of the main difficulties in this kind
of investigation. 
The picture in Fig. 1 seems to suggest an imitation
behavior. However, it is known that this behavior occurs in response
to a predator. Therefore, how can one be sure that the caterpilars
do not just react individually to the same danger signal?

\begin{figure}[htb]
\ifnum\arxiv=0 \centerline{\psfig{width=6cm,figure=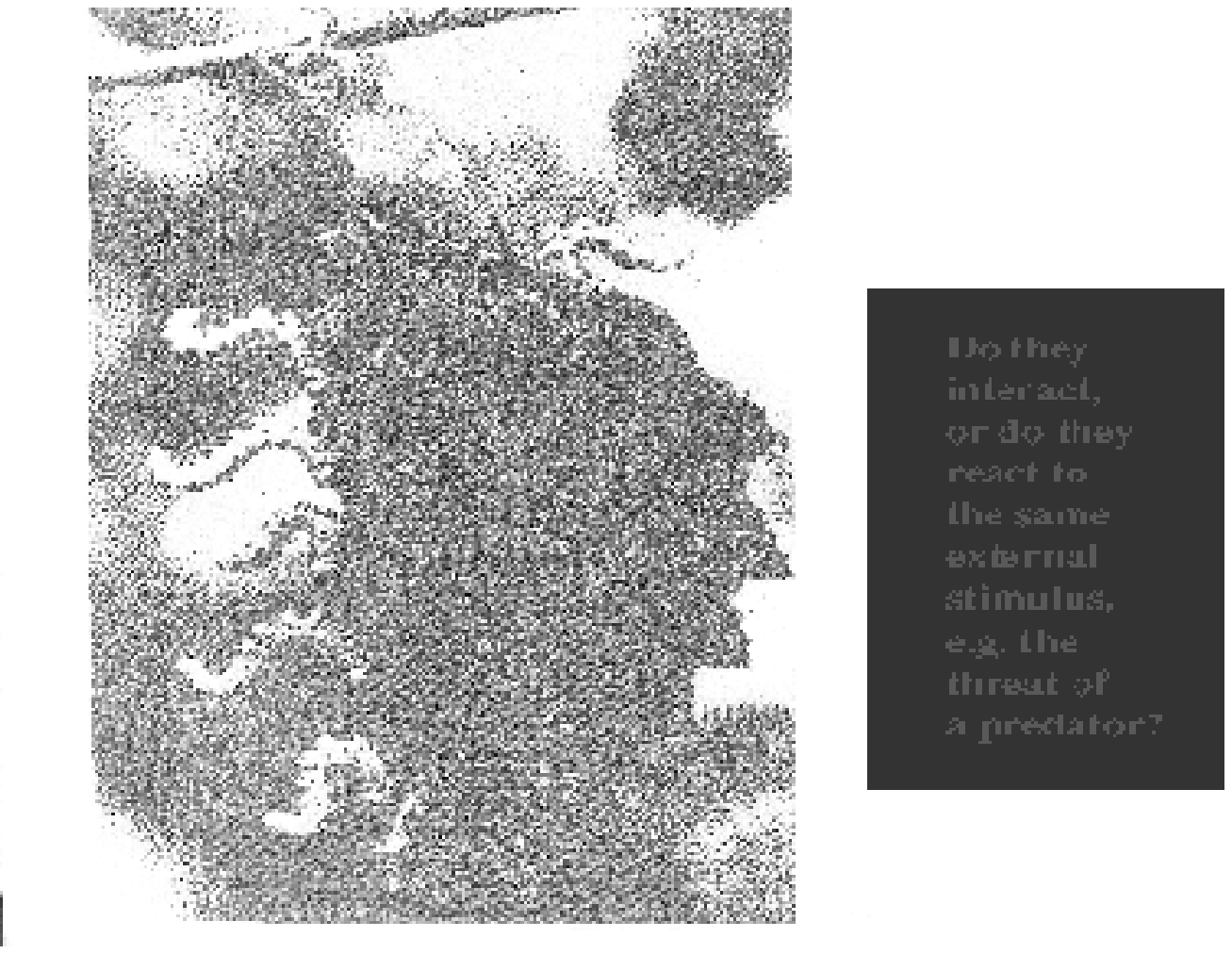}} \fi
\ifnum\arxiv=1 \centerline{\psfig{width=6cm,figure=caterpil.eps}} \fi
\qleg{Fig. 1: Synchronous movements of caterpillars.}
{The caption says that the picture ``indicates that there is an
exchange of information between individuals''. This is not
necessarily true, however. It can also be a common response
to an external stimulus. The parallel orientations of iron filings
sprinkled on a white card placed on top 
of a permanent magnet are not due to any interaction between 
them but solely to the external magnetic field.
This illustrates what is probably the main difficulty when
probing interactions.}
{Source: Costa (1997)}
\end{figure}

An interaction is characterized by its strength and range.
In biological systems it is is also characterized by its
duration. Indeed, 
in contrast to physical systems in which interactions are
permanent, systems of living organisms may have
temporary interactions. The most obvious example is the
interaction between males and females. It occurs only in mature
animals and usually follows some seasonal pattern.

\qA{Classes of collective motions}
By the term ``collective motions'' we wish to refer
to the motions of many individuals in a population, whether
they interact in some way or not. 
\qpar
Cases ranging from
complete randomness to
fairly ordered motions can be illustrated by the following examples.
\qeeb{1} 
{\color{blue} Independent motions of random individuals.}\qL
Ex: Motions of many Brownian particles in a liquid.
\qeeb{2} 
{\color{blue} Independent motions of deterministic individuals.}\qL
Ex: Movements of people on a square or the departure hall of a
railway station. All persons know where they are going but the
global picture is not very different from the former.
\qeeb{3} 
{\color{blue} Collective motion of non-interacting individuals:
hydrodynamic interaction}\qL
When microorganisms swim fairly fast in water and when in addition
their density is high enough the drift of the water induced
by the movement of one individual will affect the movements
of its neighbors. In this way, the trajectories may become correlated
even though the organisms do not have any
direct means of interaction. This effect is known as
hydrodynamic interaction.\qL
Ex: A spectacular example was studied by Rabani et al. (2013).
In this case the bacteria were moving in a two-dimensional
film of water which greatly enhanced the hydrodynamic effect.
As a result,
individual velocities were fairly parallel to one another.
\qeeb{4} 
{\color{blue} Collision avoidance and the sidewalk effect}\qL
When ants come close to one another they usually
stop for a few seconds in order to examine one another.
Swimming microorganisms such as Euglena gracilis 
\begin{figure}[htb]
\ifnum\arxiv=0 \centerline{\psfig{width=6cm,figure=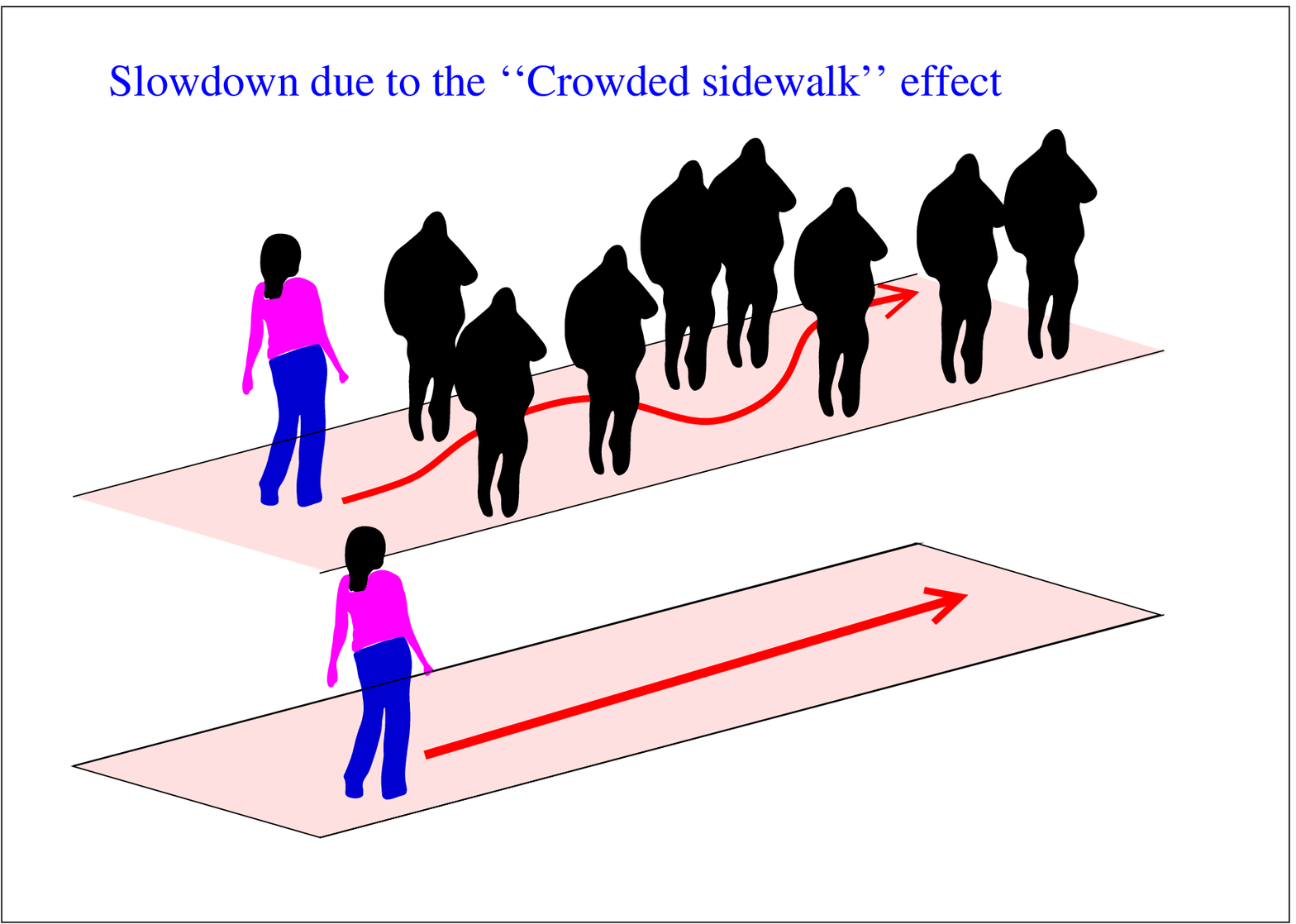}} \fi
\ifnum\arxiv=1 \centerline{\psfig{width=6cm,figure=sidewalk.eps}} \fi
\qleg{Fig. 2: Illustration of the ``crowded sidewalk''
effect.}
{In a sense collision avoidance can be 
considered as being an interaction.
It is similar to the kind of weak interaction that occurs between
the molecules of a gas except that in a gas if one assumes the
encounters to be elastic there is no slowing down effect.}
{}
\end{figure}
must of course avoid collisions  
but their interaction when they are close to one another 
seems to be very brief.
In other words, whether or not there is any ``real'' interaction
there will be a slowing down effect due to the necessity of
avoiding head-on collisions. We will see later on
that this effect can be observed statistically.
\qeeb{5} 
{\color{blue} Attraction between individuals: clustering and shoaling}\qL
Ex: A shoal of fishes consists in a population of fishes remaining
in the same location. Similarly, as we will see below,
ants or bees left to themselves move toward one another and
eventually form one big cluster. In the absence of any external
stimulus this can hardly be explained otherwise than by an
attractive force between them. 
\begin{figure}[htb]
\ifnum\arxiv=0 \centerline{\psfig{width=6cm,figure=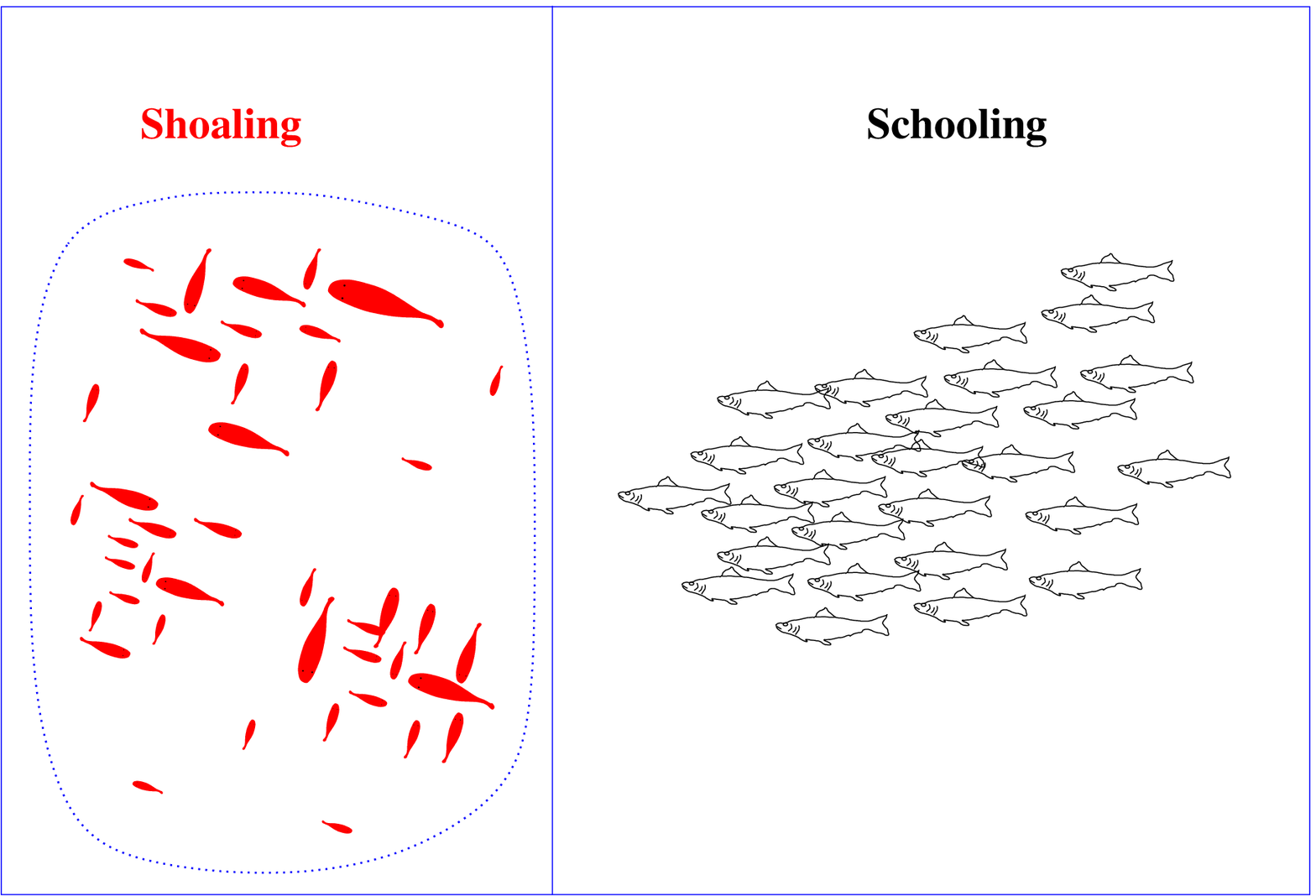}}\fi
\ifnum\arxiv=1 \centerline{\psfig{width=6cm,figure=schooling.eps}} \fi
\qleg{Fig. 3: Difference between shoaling and schooling.}
{Shoaling means that a group of individuals remains in the
same location. It is similar to clustering except for the
fact that it can be a permanent situation whereas clustering
is rather temporary. Schooling is similar but means that the group
moves together. Both shoaling and schooling reveal an
underlying inter-individual attraction. However, shoaling
may also just reflect gathering in an appropriate environment.}
{}
\end{figure}
%
\qeeb{6} 
{\color{blue} Collective motion of interdependent individuals}\qL
Ex: The V-shaped flight formation of geese or ducks or
the collective motion of a school of herrings.

\vskip 5mm
\centerline{\LARGE \bf \color{magenta}
Part 2: Examples of inter-attractive behavior}
\vskip 3mm

In this part we focus on cases for which there is clear evidence
of collective behavior brought about by inter-individual 
attraction forces. 

\qI{Clustering as a means for estimating inter-attraction}

The occurrence of clustering in a population {\it in the
absence of any external stimulus} (whether light, food, smell
or any other) provides good evidence for the 
existence of an attraction force, or more precisely
it shows that the attraction force is strong enough to
overcome the dispersion effect of individual 
and more or less random velocities. 
in what follows this dispersion effect will be referred to 
simply as noise.
\qpar
The physical analog of clustering
is the phenomenon of condensation
(i.e. transition from a gas to a liquid) which is 
a (first order) phase transition. In this case, the noise 
is due to what is called thermal agitation. 
As the van der Waals interaction
forces between molecules are basically the same whether 
the material is in gas or liquid state,
condensation occurs when a reduction in temperature makes
the noise $ kT $ 
become smaller than the attraction effect. This is a self-reinforcing
effect because by reducing inter-molecular distances condensation
at the same time increases interaction strength. 
\qpar

Observation of clustering provides 
in a fairly easy way an insight into the attractive
forces between living organisms. That is why in the next section
we will describe several cases of clustering.
One should keep in mind that whereas clustering is evidence
of attraction, the opposite is not true. Attraction
without clustering simply means that the dispersion forces
are too strong.

\qI{Attraction strength among social insects: ants and bees}

First, we will study the clustering of  
ants and bees, i.e. two social insects.
Then, we show that some non-social insects also form clusters.
Naturalists call them gregarious insects. 
Finally, we 
describe the case of drosophila for which no
clustering  is observed. They are sometimes called
solitary insects, but it is clear that even such 
insects have interactions, among which the 
attraction between males and females is the most obvious.

\qA{Ants}

Fig. 4a shows that the behavior of ants is the exact
opposite of a gas. Whereas a gas tends to occupy all
the volume available, ants will cluster into
one half of the container. As a matter of fact, they
will not even occupy the totality of the right-hand
half of the container but only a small part of it.
\qpar
\begin{figure}[htb]
\ifnum\arxiv=0 \centerline{\psfig{width=8cm,figure=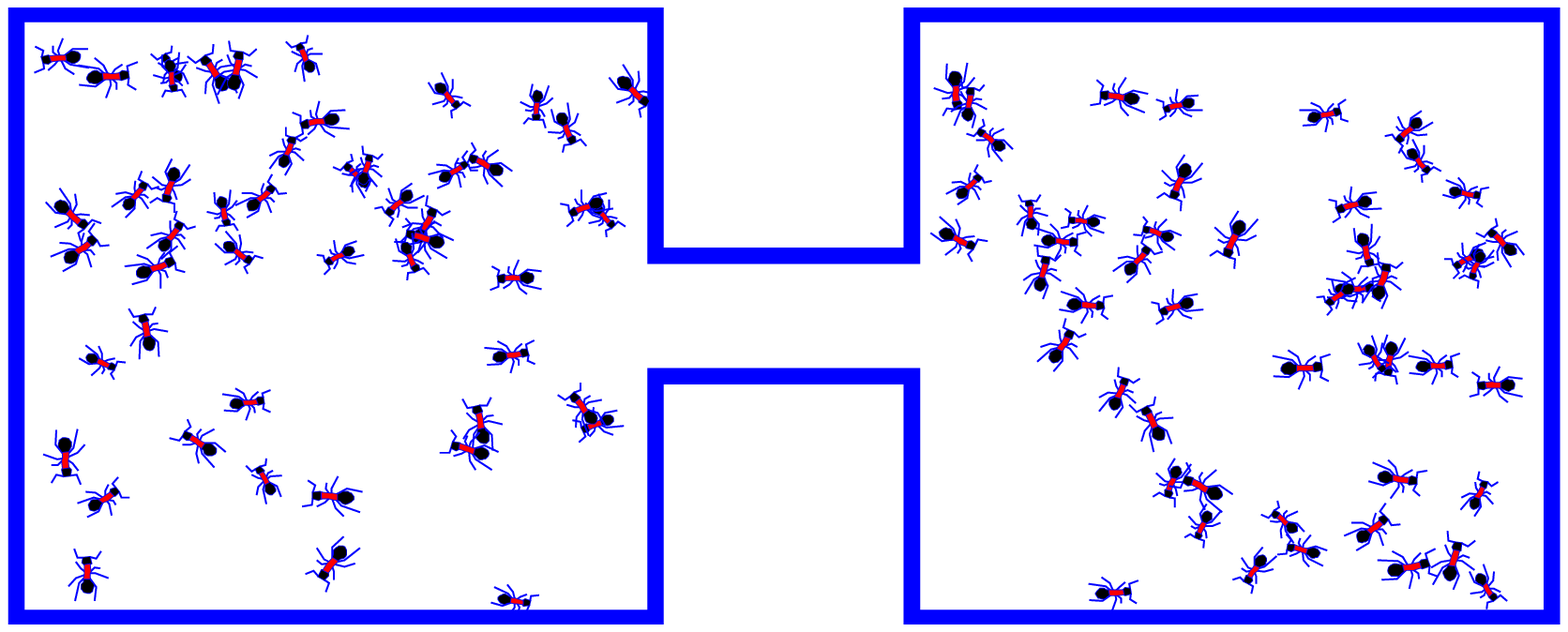}}\fi
\ifnum\arxiv=1 \centerline{\psfig{width=8cm,figure=ant1.eps}}\fi
\vskip 2mm
\ifnum\arxiv=0 \centerline{\psfig{width=8cm,figure=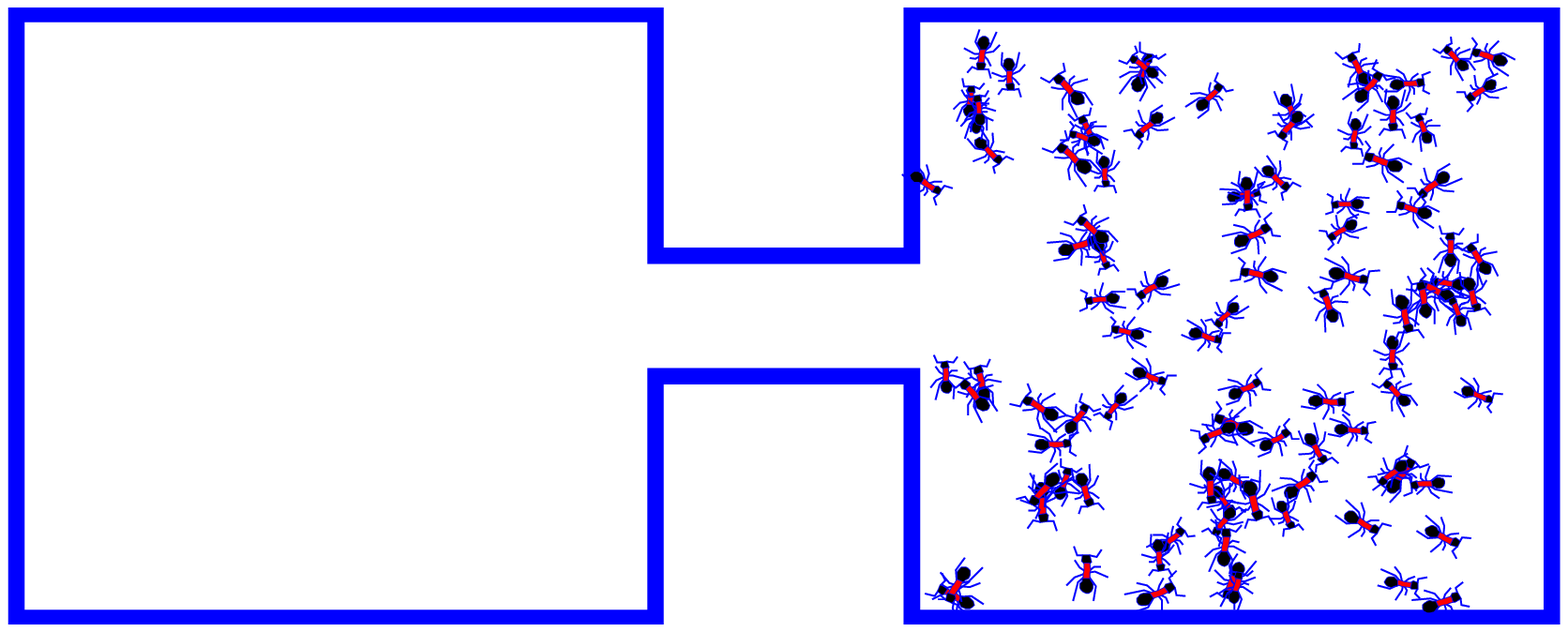}} \fi
\ifnum\arxiv=1 \centerline{\psfig{width=8cm,figure=ant2.eps}} \fi
\vskip 1mm
\qleg{Fig. 4a: Clustering of red fire ants.}
{The ants move from
an initial state where
they are on separate sides to a state where almost
all of them are on the same side. 
The boxes had an area of
about 60 square centimeters. The connecting glass tube had a length of
3 centimeters and a diameter of one centimeter.
The experiment was done
in August 2011 at room temperature that is to say at about
25 degree Celsius. 
The initial numbers $ N $ of ants on each side were 100, 200, 300, 500,
750, 1000. For each $ N $ the experiment was repeated 5 times.
For these repetitions the coefficient of variation was on average 65\%.
Contrary to what happens for a gas
the flow velocity
does {\it not} increase when the
density becomes higher. On the contrary it decreases.
This means that the process is more and more dominated by
attraction. 
\qL
Just to give an order of
magnitude of average traffic in the communication
tube, it can be observed that for an initial population
of 1,000 on each side the clustering process lasted 6.8 hours
which means that
the average flow of ants per minute was
$ 1000/(6.8\times 60)=2.4 $. This shows that the section
of the communication tube was not in itself a limiting factor.}
{Source: The data are from an experiment done in August 2011
by Dr. Lei Wang
at the South China Agricultural University in Guangzhou
(personal communication).}
\end{figure}

\begin{figure}[htb]
\ifnum\arxiv=0 \centerline{\psfig{width=10cm,figure=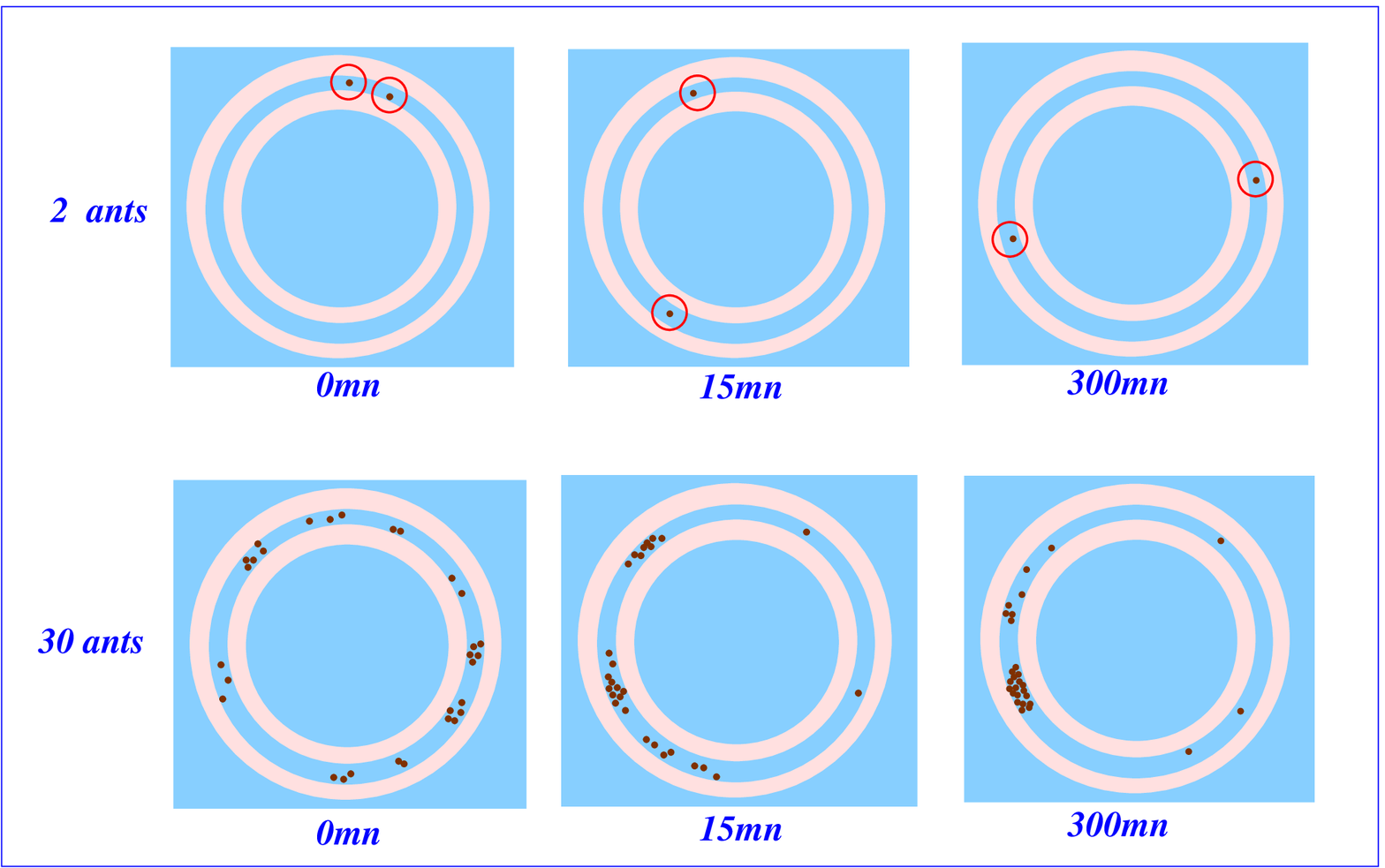}}\fi
\ifnum\arxiv=1 \centerline{\psfig{width=10cm,figure=ring6.eps}}\fi
\qleg{Fig. 4b: Clustering process for ants.}
{The ants belong to the species {\it Terramorium, Caespitum}, 
Linnaeus 1758; their length is about 1-2mm;
they were bought on Taobao, a well-known
Chinese shopping website.
The ring structure (made of polymer clay and covered by a plastic plate)
prevents
the ants from remaining motionless at the edge of the container
as would happen with an ordinary container. The ability to form
a cluster (as well as the speed of the aggregation process)
depends upon the number of ants. The same observation holds for
bees (see Table 1).
Two ants on average remain fairly apart
as shown in the top panel. On the contrary, for a group of 30 ants
after a 
time interval of the order of one hour they form a cluster
(or sometimes two clusters).
In this experiment
the diameter of the ring was 8cm.}
{Source: The experiment was done in March 2014 
by Li Geng at the
School of Systems Science of Beijing Normal University.}
\end{figure}
In the experiment of Fig. 4b
the ``edge difficulty'', namely the fact that the ants tend
to go to the edge of the container and
stay there without moving at all
was solved through the ring structure. 
In the experiment of Fig. 4c it was solved in a different way.
The rectangular space in which the ants
were spread was surrounded by 4 strips of paper
which had been recovered with a repellent liquid 
whose smell the ants do not like. Before being used
the paper strips were left to dry but the smell was
still perceptible which means that they acted as 
a repellent%
\qfoot{However, when the space allocated to the ants
was small, a fraction of them happened to cross the
border. This occurred particularly at the start
of the experiment when the ants were excited after
having been removed from the rest of the colony.}%
.
\qpar
\begin{figure}[htb]
\ifnum\arxiv=0 \centerline{\psfig{width=14cm,figure=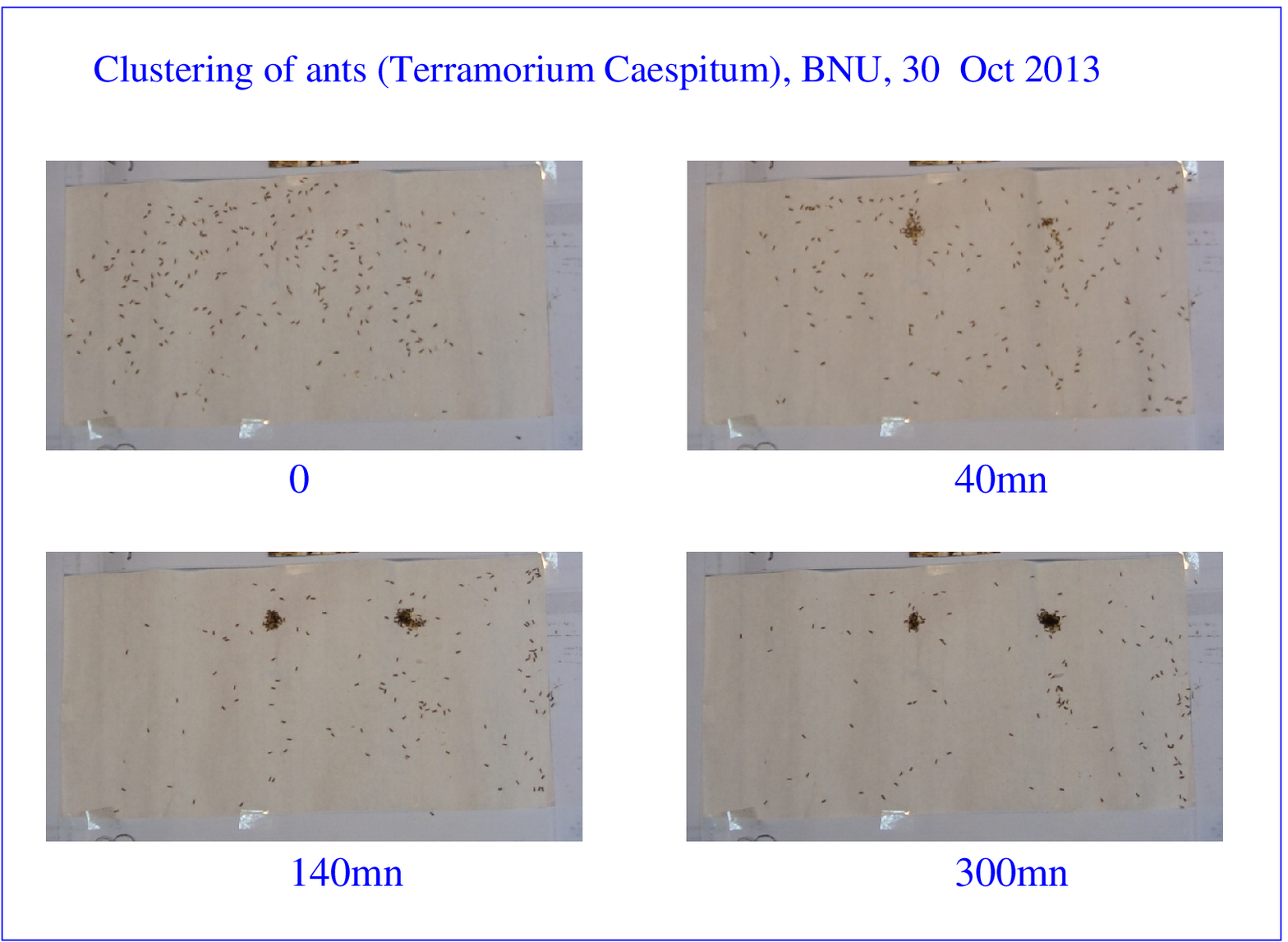}}\fi
\ifnum\arxiv=1 \centerline{\psfig{width=14cm,figure=antclus.eps}}\fi
\qleg{Fig. 4c: Clustering process for ants.}
{Strips of paper impregnated with a repellent
kept the ants away from the borders of the area.
In this experiment there were 229 ants (density=57/sq.decimeter). 
Similar experiments were done with
different densities and number of ants. 
An  approximate rule of thumb that emerged is 
that the larger the  number and density
the less time it takes for big clusters to appear and the
greater the proportion of ants which are in the cluster 
(see Fig. 4d).}
{Source: The experiment was done on 30 October 2013
and March 2014 at the
School of Systems Science of Beijing Normal University.}
\end{figure}

Two qualitative rules emerge from these
observations.
\qbu The higher the density, the faster the aggregation process.
\qbu When the density was low the cluster which appeared
at long last comprised only a small fraction of the ants. 
\qL
The graph of Fig. 4d documents the second effect.
Observation shows that the dispersion (i.e. the proportion 
of ants not in the cluster) is higher in two dimensions
than in one dimension, i.e. in the ring experiment. 
%
\begin{figure}[htb]
\ifnum\arxiv=0\centerline{\psfig{width=8cm,figure=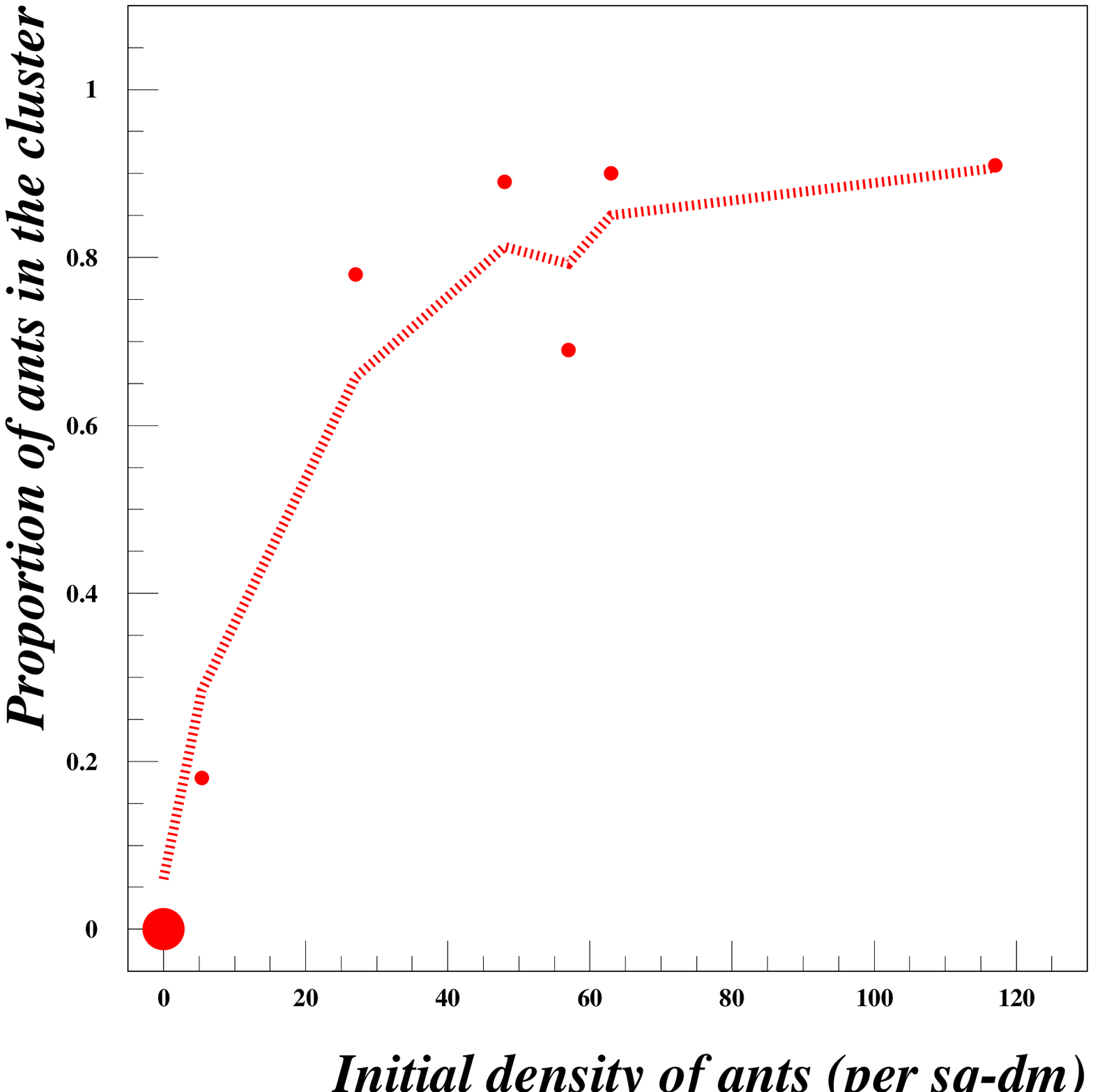}}\fi
\ifnum\arxiv=1 \centerline{\psfig{width=8cm,figure=densprop.eps}}\fi
\qleg{Fig. 4d: Relationship between density and the proportion
of ants in the cluster.}
{The small red dots are for experiments similar to those
shown in Fig. 4c. A big dot was drawn at the origin 
because one can be certain that when the density tends
toward zero the ants will be unable to find one another
which means that no aggregation can occur. For the sake
of simplicity }
{Source: The experiments were done in Oct-Nov 2013
at the
School of Systems Science of Beijing Normal University .}
\end{figure}

\qA{Bees}

Beekeepers know very well that bees 
cluster together under some special circumstances. 
\qbu In wintertime, when the temperature is low, they form a cluster%
\qfoot{It is often said that it is for raising their temperature
at least for those inside the cluster. 
However, it is also known by beekeepers that
bee colonies start to become active in mid-February (in
countries whose climate is similar to that of France)
that is to say at a moment when the temperature outside
of the beehive is still fairly low. It is in May and June
that the production of honey is usually highest, not
in July or August.}
\qbu In the process of swarming, that is to say 
when the ``old'' queen leaves with {\it part} of the
colony or in the similar process of
absconding, that is to say when the queen leaves 
with {\it all} the colony.
\qpar
The fact that bees form clusters {\it without any
external stimuli} was discovered by J. Lecomte 
in the late 1940s. He explored carefully the conditions under
which clustering takes place and published his
results in three papers (1949,1950,1956) which, although 
an important contribution to the understanding of the
{\it collective} behavior of bees, were completely 
overlooked and forgotten in the following decades.

\begin{figure}[htb]
\ifnum\arxiv=0 \centerline{\psfig{width=15cm,figure=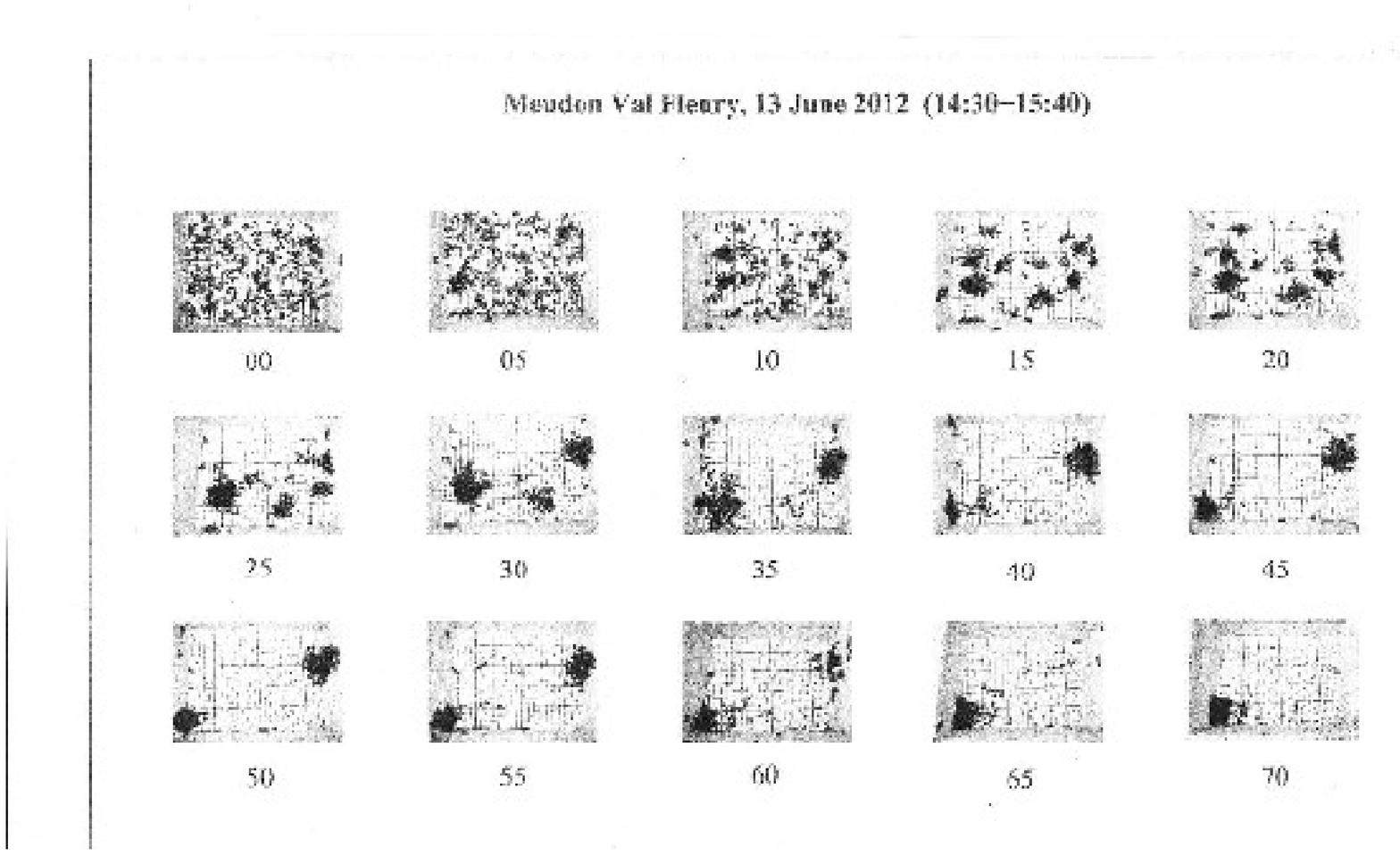}}\fi
\ifnum\arxiv=1 \centerline{\psfig{width=15cm,figure=bee1.eps}}\fi
\qleg{Fig. 5a: Clustering process for bees 
(Apis mellifera mellifera).}
{Altogether there were about 300 bees. Initially they were put to
sleep through 5 minutes in carbon dioxide. All the bees are
from the same beehive. The numbers under the pictures give
the time in minutes.}
{Source: The data are from an experiment done in Meudon
Val Fleury near Paris
in July 2012 by Jack Darley and Bertrand Roehner.}
\end{figure}

Fig. 5b shows that clustering also occurs in a mixed
population of bees coming from two different colonies. 
\qpar
This result is not completely unexpected because beekeepers
know that it is possible to put frames from different colonies 
into the same beehive. Fights may be prevented 
by spraying all the bees with flavored water.
Nevertheless, it is also true that when a bee 
tries to enter into the beehive of a different colony 
is will be identified and chased away by the bees which stand 
guard next to the entrance%
\qfoot{Does this process occur
systematically, that is to say
with a probability of one, or rather with a probability $ p<1 $,
which would imply that some foreign bees can get entrance?
This point is not clear from what we have read so far.}%
.
%
\begin{figure}[htb]
\ifnum\arxiv=0
\centerline{\psfig{width=5cm,angle=90,figure=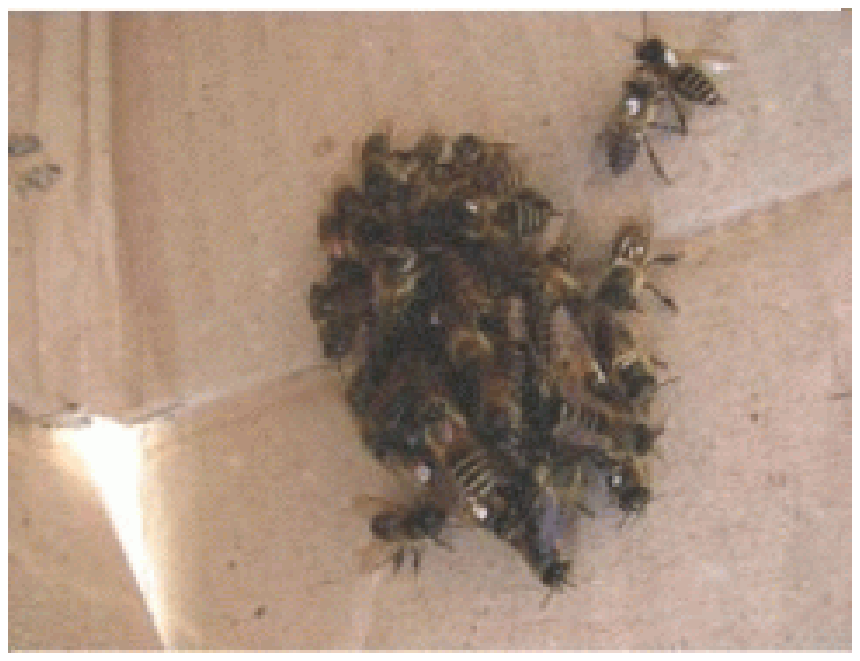}}\fi
\ifnum\arxiv=1
\centerline{\psfig{width=5cm,angle=90,figure=mixedcluster.eps}}\fi
\vskip 2mm
\qleg{Fig. 5b: Mixed cluster of bees.} 
{It is often said that bees from different colonies
do not mix with one another or may even fight one another.
However, this experiment shows a case where they
form  mixed clusters.
The bees from colony $ A $ are marked with a white dot whereas
those of colony $ B $ have no dot. The picture
shows that the cluster comprises
bees from the two colonies.}
{Source: The experiment was performed in December 2011
by Zengru Di, Bertrand Roehner, Ken Tan and Zhengwei Wang
at the Eastern Bee Institute, 
Yunnan Agricultural University, Kunming, China.}
\end{figure}

The observations summarized in Table 1 suggest that 
for the formation of a single cluster there is a
critical minimal population threshold which is of the order
of 40 bees. 
%
\begin{table}[htb]

\centerline{\bf Table 1\quad Influence of the number 
(and density) of bees on cluster formation}

\vskip 2mm

\hrule
\vskip 0.5mm
\hrule

\vskip 2mm
$$ \matrix{
\qtb
 \hbox{Number of bees} \hfill & 5 & 10 & 15 & 25 & 50 & 75 \cr
\noalign{\hrule}
\qth \qtb
\hbox{Frequency of formation of a single cluster}\hfill &
                     30\% & 40\% & 6\% & 6\% & 80\% & 100\% \cr
\noalign{\hrule}
} $$
\vskip 2mm
{
Notes: The experiments were performed at 25 degree Celsius.
For each total number of bees the experiment was repeated 18 times.
The same box was used in all experiments which means that
the density of the bees per square centimeter decreased along with the
number of bees. A cluster was defined as an aggregate
containing at least 80\% of the 
total number of bees. \qL
The results show a fairly sharp
transition between 25 and 50 bees. \qL
There
is no clear explanation for the fact that the probability
increases again for 5 and 10 bees; of course, for such small numbers
the variability may be large which means that 
in addition to the average, one would also
need to know the standard deviation.
\qL
{\it Source: Lecomte (1956).}}
\vskip 2mm
\vfill

\hrule
\vskip 0.5mm
\hrule
\end{table}
%

\qI{Attraction strength among non-social insects}

\qA{Comparative experiment}
First we describe an experiment that was done in parallel with drosophila
and with beetles (Alphitobius diaperinus).
\qpar
One takes a test tube containing some 50
drosophila and one makes them all move to the bottom of the tube
by hitting the tube on a table. Then, very quickly%
\qfoot{This movement must be fast because drosophila have a
natural tendency to go upward.}
one puts the tube
on the table in horizontal position. Let us assume that the bottom
of the tube is on the right.
After a few seconds, some 5 flies will have reached the
left-hand side, and may be 10 others will be in the middle of the tube.
If one waits 5mn, the flies will be distributed fairly uniformly
throughout the tube. 
\qpar
If one repeats the same experiment with beetles
it will be seen that after 5mn almost all insects are still 
together on the right-hand side of the tube.
\qpar
A physical interpretation of this experiment can be proposed.\qL
Suppose that we replace the drosophila by the molecules of a
gas and the beetles by the molecules of a liquid.
By hitting the tube on the table all molecules will move
to the bottom%
\qfoot{Because the velocity of the tube 
must be higher than the velocity of the molecules,
instead of a test tube the experiment
would require a steel cylinder! This 
is rather a thought experiment.}%
.
For this experiment we do not need to put the
tube in horizontal position; this was only required because
of the tendency of drosophila to go upward.
After the shock, the gas molecules will
re-occupy the whole available room of the test tube
within a fraction of a second.\qL
In the case of a liquid, it
will go to the bottom of the tube by the mere
effect of gravity. Once there, only a small proportion of the
molecules will occupy the volume of the test tube above the liquid.
This fraction corresponds to the so-called equilibrium vapor
pressure. 
\qpar
For instance, for water at 20 degree Celsius, the
equilibrium vapor pressure $ p $ is 2.3 kilo-Pascal. 
For our purpose,
the pressure is of little significance. One would rather wish
to know the density $ \rho $ which corresponds to $ p $. Fortunately,
if one assumes that the vapor above the liquid can be described
as an ideal gas (which is certainly true at such a low pressure)
the two variables are related by the equation of state of
an ideal gas which can be written:
$$ p=\rho r_wT, \hbox{ where } T=\hbox{Kelvin temperature, } r_w=R/M_w $$

$ R $ is the gas constant: 
$ R=8.3\ \hbox{kg} \hbox{m}^2\hbox{s}^{-2}\hbox{K}^{-1}$
and $ M_w $ is the mass of one mole of water: $ M_w=18 $g.
Thus, $ r_w=462\ \hbox{m}^2\hbox{s}^{-2}\hbox{K}^{-1}  $.
For the density one gets: 
$$ \rho=p/r_wT=2300/(462\times 293)=0.017\hbox{kg/cubic-meter}=
17\hbox{mg/liter} $$
If we remember that the density of air in standard room conditions
is 1,200mg/liter, we see that the density of water vapor is
about 100 times smaller. 
\qpar
When the inter-molecular attraction
is smaller the boiling point will be lower and the
vapor pressure will be higher. For instance, at a temperature
of 20 degree Celsius, acetone has an equilibrium vapor pressure
of 22.8 kPa. In addition, $ r_a=r_w/3.2 $. As a result, the density
of acetone vapor will be about 32 times higher than for water.
In contrast, the ratio of the boiling temperatures of water and
acetone is only: $ 100/56=1.8 $. 

%
\begin{figure}[htb]
\ifnum\arxiv=0\centerline{\psfig{width=7cm,figure=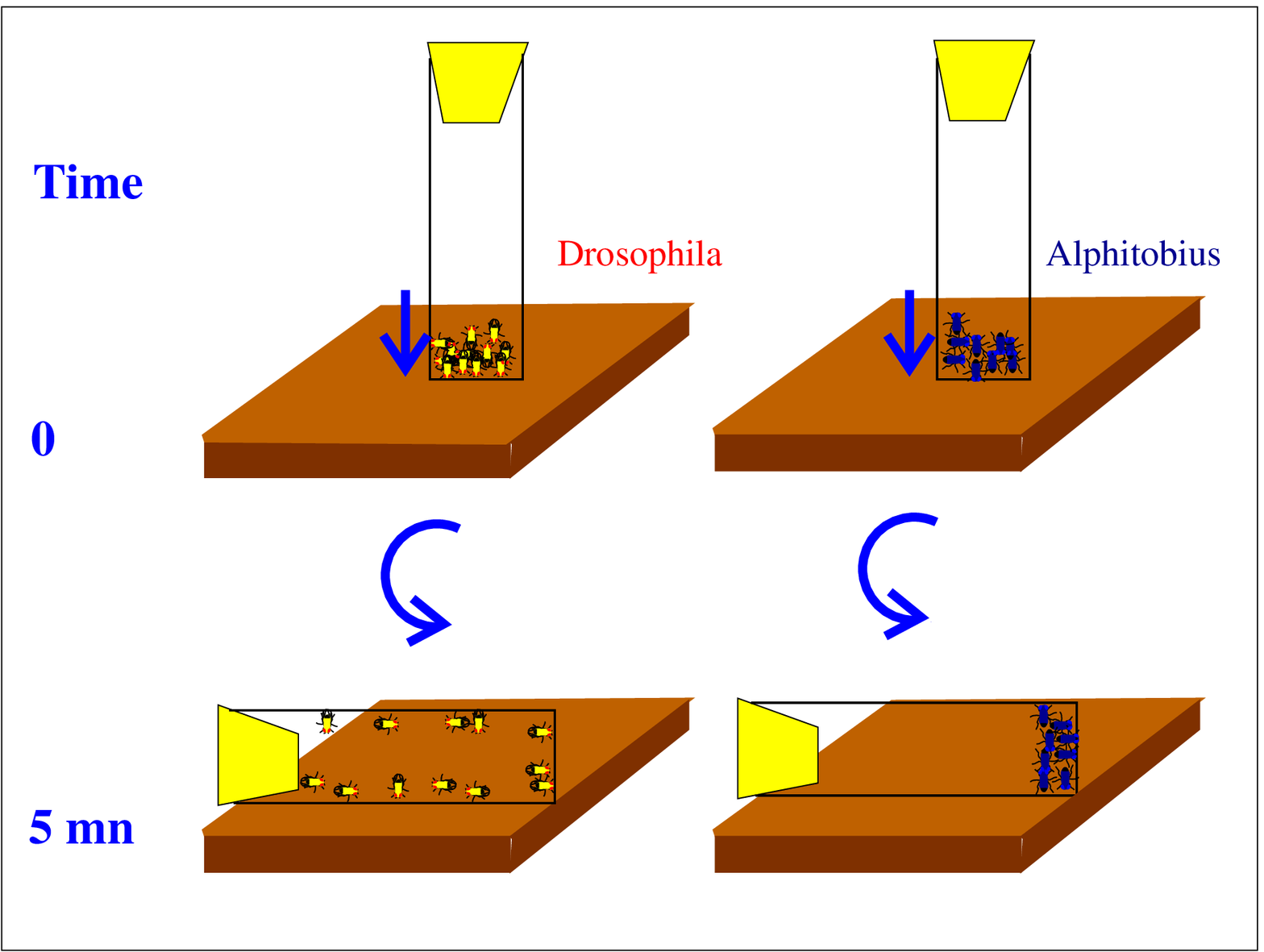}}\fi
\ifnum\arxiv=1\centerline{\psfig{width=7cm,figure=expdrobee.eps}}\fi
\qleg{Fig.\qhu 6\qhv Experiment demonstrating the existence
of dispersion forces.}
{This test-experiment which takes only a few minutes shows that 
there is a striking contrast between the behavior of drosophila
and that of beetles (Alphitobius diaperinus).
It could possibly be argued
that the beetles do not move just because they are not active.
In other words the fact that they remain together does not 
in itself prove the existence of attraction forces. However,
it is difficult to explain the behavior of the drosophila
without assuming the existence of agitation forces which are not
kept in check by inter-individual attraction.}
{Source: Lai Shu Ying and Feng Meng Ying, Report for the ``Championship
in Experimental Science'' at Beijing Normal University (Dec 2012).} 
\end{figure}

\qA{Beetles}

The difference in behavior encapsulated in the previous experiment
needs to be studied in more detailed way in separate 
clustering experiments. Fig. 7 summarizes an experiment for 
beetles. 
%
\begin{figure}[htb]
\ifnum\arxiv=0
\centerline{\psfig{width=12cm,figure=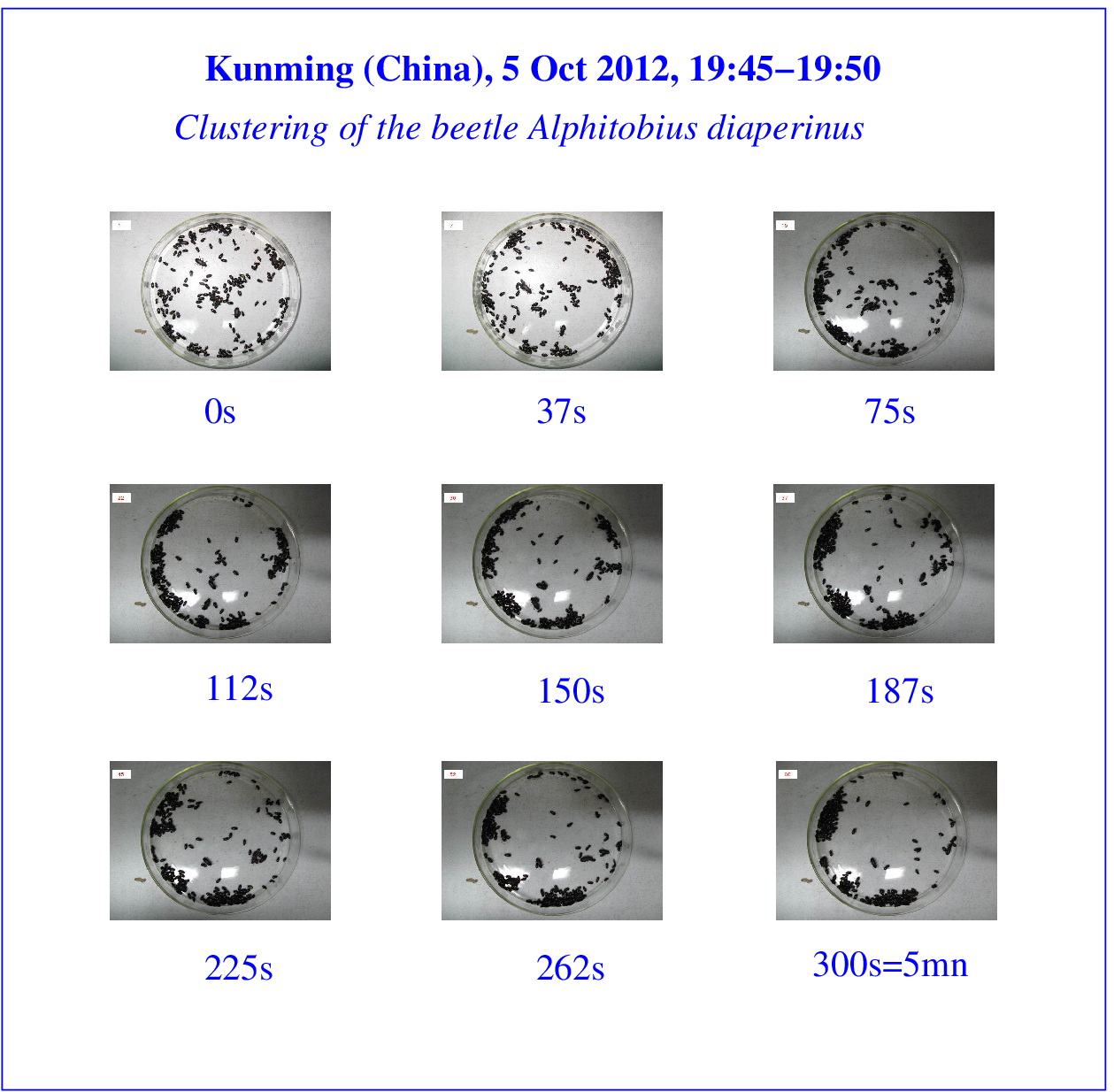}}\fi
\ifnum\arxiv=1 \centerline{\psfig{width=12cm,figure=beetle.eps}}\fi
\qleg{Fig.\qhu 7\qhv Experiment showing the aggregation 
of beetles (Alphitobius diaperinus).}
{The first action of the beetles is to rush toward the edge
of the container. After this phase which lasts about 75s
the positions of the beetles change only slightly. The diameter
of the container was 15.5 cm.}
{Source: The experiment was performed on 5 October 2012
by Bertrand Roehner and Zhengwei Wang
at the Eastern Bee Institute, 
Yunnan Agricultural University, Kunming, China.} 
\end{figure}

It should be noted that because these beetles prefer dark places
the experiment must be done under fairly uniform light.
\qpar
In the future, 
such experiments should be repeated with a larger number of
beetles and possibly in a ring-shaped container in order
to eliminate the edge effect.
\qpar
These beetles are the adult form of the so-called buffalo worms
that are sold as foodstuff for birds or fishes.
The transition to the adult form seems to be very dependent upon
temperature and air humidity. At 20 C (and fairly dry air) it may
take 3 months, whereas around 30 C it may take less than one month.

\qA{Drosophila}
Drosophila are used extensively in genetics
laboratories which makes them easily available. 
They  have a natural tendency to climb upward. Thus, 
in the laboratory tubes in which they are kept, 
they are usually concentrated on the food which fills
the bottom of the tube and at the top under the 
cap of the tube.
\qpar
In the experiment described below, the 
Plexiglas containers had a height
of only 6 mm which prevented from flying. We used low intensity,
``cold'' light which did not increase the temperature of
the container. under these conditions we did not observe 
any clustering.
Of course, it is quite possible that under different conditions
an aggregation process may occur.

%
\begin{figure}[htb]
\ifnum\arxiv=0
\centerline{\psfig{width=14cm,figure=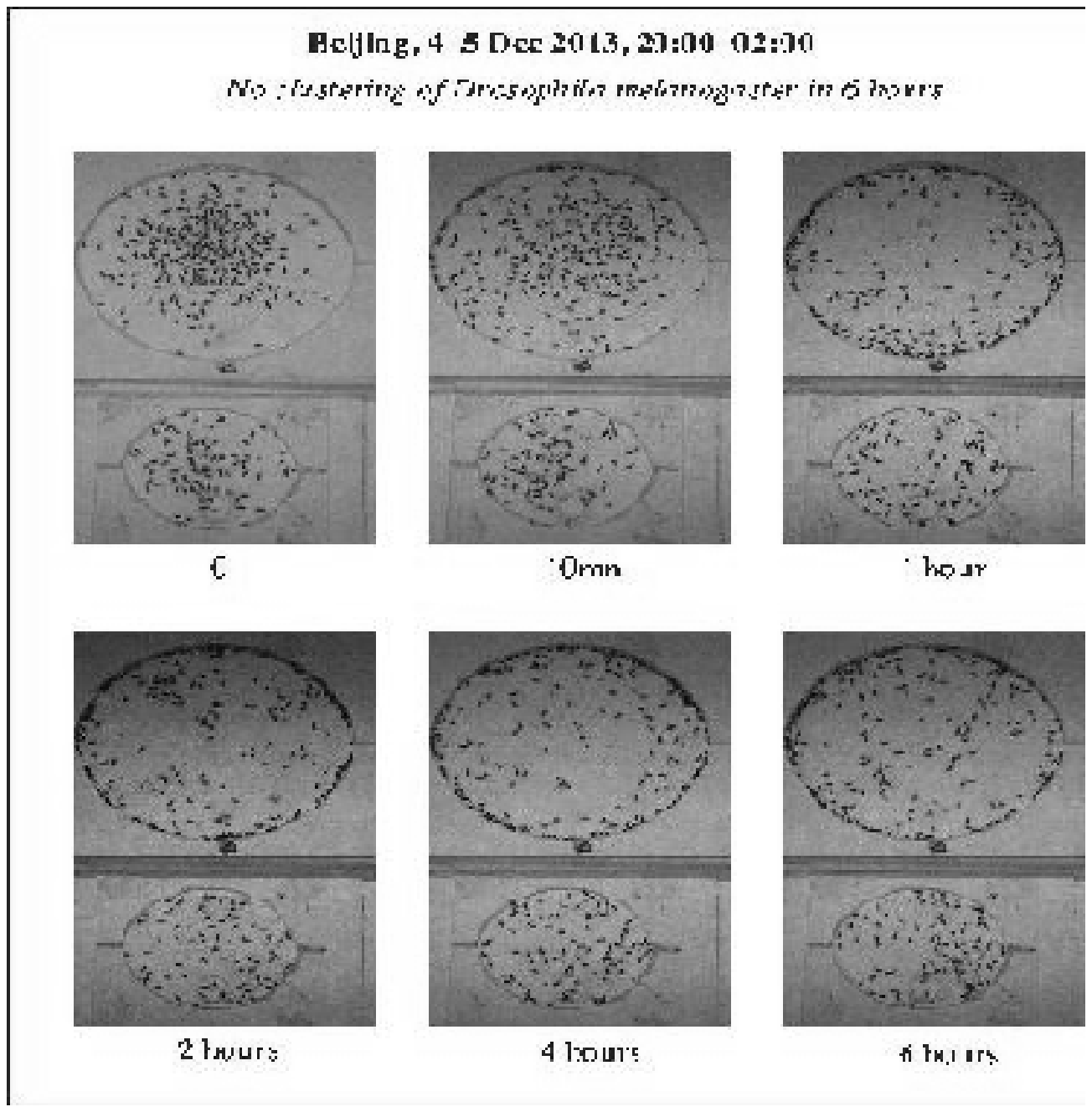}}\fi
\ifnum\arxiv=1 \centerline{\psfig{width=14cm,figure=drosocl.eps}}\fi
\qleg{Fig.\qhu 8\qhv Experiment showing that drosophila do 
not cluster together.}
{In the lower cell there were 95 flies (density=247/sq-dm),
whereas in the upper cell there were 244 flies (density=
309/sq-dm). At the end, after 6 hours, the density is fairly
uniform in the lower cell; in the upper shell it is higher
on the edge than in the middle but there is no
clustering location along the edge. Of course, this does
not mean that there are no interactions whatsoever among
fruit flies. For instance, there is an obvious attraction
between males and females.}
{Source: The experiment was done on 4-5 December 2013
at the School of Systems Science of
Beijing Normal University. The fruit flies came from
the genetics laboratory of Pr. Dou Wei (BNU). 
Many thanks to him.} 
\end{figure}

\vskip 5mm
\centerline{\LARGE \bf \color{magenta}
Part 3: Lessons and hints from physical chemistry}
\vskip 3mm

\qI{How to use aggregation/diffusion to measure interaction?}

In this section we wish to examine what can be learned from
physics about possible methods based on aggregation
or diffusion for measuring interactions. This is
a very natural step because physics has a long record
of studying interactions between particles.

\qA{Aggregation}

At first sight, aggregation phenomena may seem at odds with
what one is used to observe in physics. Indeed, diffusion
seems much more common than aggregation or clustering. 
For instance, how can one get an aggregation effect similar
to the one described in  Fig. 4a? \qL
Consider  a mixture of oil and balsamic vinegar. As the two liquids
are not miscible and the vinegar has a higher density it will 
sit at the bottom of the bottle. Shaking the
bottle  will create a suspension of small, brown droplets of
vinegar within the oil. However, within a few minutes the
two liquids will separate again. This process
is of course facilitated by the density difference 
but it also occurs without it as shown by the
aggregation process of patches of oil floating on water 
or vinegar.
\qpar

Aggregation will occur for systems characterized by a strong
attraction and a low level of noise. In all other
cases there will be diffusion, but the diffusion can
be fast or slow,
This leads to question
the effect of inter-molecular attraction on
diffusion? Can physics give us a method for deriving
attraction strength from diffusion data?

\qA{Connection between diffusion and attraction}

Surprisingly, when one
tries key-words such as ``diffusion''+``attraction'' on an Internet
search engine, one gets very few significant results.
The mathematical theory of diffusion which leads to the 
partial differential equation known as the diffusion equation
is purely based on random movements of {\it independent} particles.
It describes well the diffusion of Brownian particles that
is to say particles which are much bigger than molecules.
However, it is not appropriate for the diffusion of
molecules $ A $ (e.g. dye molecules) among molecules $ B $
(e.g. water molecules) for in this case the $ A-A $ and
$ A-B $ interactions play an essential role. This fact is
made obvious by observing that depending upon these
interactions, the  $ A $ and $ B $ liquids will be completely
miscible (e.g. water and ethanol), partially miscible,
or almost not miscible (e.g. water and oil). Clearly,
the degree of miscibility will affect the diffusion process. 
\qpar
Coming back to our previous Internet search,
in order to improve the results, one should rather use the
expressions ``self-diffusion'', also called
``tracer diffusion'', and ``collective diffusion'', also called
``mutual diffusion''%
\qfoot{In this connection one can mention the following 
papers: Phillies (1974), Phillies et
al. (1976), Van den Broek et al. (1981), Holmberg et al. (2011).}%
. 
\qL
What physical phenomena do these expressions represent?
\qbu Self-diffusion refers to the diffusion of individual
molecules 
taking place in the absence of any concentration gradient;
it probes the particle-particle correlation at small distance.
In physical systems, self-diffusion simply reflects
thermal molecular agitation. In systems of living organisms
it reflects their movements. Such movements can also be seen
as a form of thermal agitation because the chemical reactions
on which motion relies will become slower if the temperature falls.
\qbu Collective diffusion refers to the diffusion of a large
number of particles (most often within a solvent) 
under the influence of a concentration gradient. This
diffusion probes the correlations at large distances. \qL
When there is no interaction between the particles, the
diffusion coefficient is independent of particle
concentration. On the contrary, for an attractive interaction
between particles the diffusion coefficient tends to decrease
as concentration increases. Thus, diffusion can be used to 
probe the strength of interactions.
\qpar

When living organisms diffuse by swimming more or less
randomly in water the only interaction that one needs 
to consider is their endogenous interaction. In other words,
the physical analog that we need to consider 
is the diffusion of molecules $ A $
(the living organisms) among molecules $ B $ (the water molecules)
when there is almost no $ A-B $ interaction.
\qpar
Two physical cases fulfill these conditions:
\qee{1} Diffusion through evaporation of the molecules of liquid 
$ A $ in air.
\qee{2} Diffusion of the molecules of liquid $ A $ in a liquid $ B $
with which they have no interaction. The lack of interaction
means that the molecules cannot bind with one another 
which implies that the liquids are not miscible.
However, even for two liquids which are not miscible, the
mutual solubilities of $ A $ in $ B $ and $ B $ in $ A $
are not zero. For the case of water and liquid alkanes
they are given in Table 2a,b along with the partial
(equilibrium) vapor pressure in air. 
\qpar
The two processes of evaporation and diffusion rely on
a the same mechanism; it is schematically described in Fig. 9.

\begin{table}[htb]

\centerline{\bf  Table 2a\ Vapor pressure and solubility in
water of liquid alkanes (at 20 degree C) }

\vskip 5mm
\hrule
\vskip 0.5mm
\hrule
\vskip 2mm

\color{black} 

$$ \matrix{
\hbox{Name} \hfill  &
\hbox{Formula} & \hbox{Solubility of alkane}& \hbox{Vapor pressure} 
 & \hbox{Density of alkane vapor}\cr
\hbox{} \hfill  &
\hbox{} & \hbox{in water}& \hbox{of alkane in air} 
 & \hbox{in air}\cr
\qtb
\hbox{} \hfill  &
\hbox{} & \hbox{[mg per liter of water]}& \hbox{[kPa]}
& \hbox{[mg per liter]}\cr
\noalign{\hrule}
\qth
\hbox{n-pentane} \hfill  & \hbox{C}_{5}\hbox{H}_{12} & 40 & 58 & 1715 \cr
\hbox{n-hexane} \hfill  & \hbox{C}_{6}\hbox{H}_{14} & 11 & 17 & 600\cr
\hbox{n-heptane} \hfill  & \hbox{C}_{7}\hbox{H}_{16} & 2& 5.3& 217\cr
\hbox{n-octane} \hfill  & \hbox{C}_{8}\hbox{H}_{18} & 1 & 1.5& 70\cr
\qtb
\hbox{n-nonane} \hfill  & \hbox{C}_{9}\hbox{H}_{20} & 0.2 & 0.5& 26\cr
\noalign{\hrule}
} $$
\vskip 0.5mm
Notes: Vapor pressure ($ x $) and solubility in water ($ y $) 
decrease almost
at the same rate. This suggests that the mechanisms of the
two phenomena may be fairly similar. The $ (\log x,\log y) $ correlation
is 0.993 and the regression leads to the relationship: \qL
\hfil $ \hbox{solubility (mg/l)}=
0.5\left[\hbox{vapor pressure (kPa)}\right]^{1.1} $ \hfil \qL
This formula can be used to predict the solubility of
decane ($ n=10 $), undecane ($ n=11 $), dodecane ($ n=12 $).
One gets (in micro-g/l): 66, 22, 3.8. As these solubilities 
are very low they are 
difficult to measure with acceptable accuracy. For decane
5 measurements performed by different authors range 
from 7 to 50 and their average is 32 micro-g/liter 
(Economou et al. 1997 p. 539). 
\qpar
The alkane solubility in water measures the diffusion of
alkane molecules
in water. Similarly, the density of the alkane vapor
measures the diffusion of the alkane molecules
in air. On average the alkane diffusion in air 
is about 80 times larger than its diffusion in water; more
precisely the ratio increases from 43 in the case of pentane
to 130 for nonane.
These ratios certainly reflect the fact that it is
easier for escaping alkane molecules to open their way
in air than through water molecules. One is not surprised
by the fact that this effect becomes stronger for bigger
molecules. 
\qpar
{\it Sources: Yang (2011, p.60). Wikipedia articles (in German).
Comparison of data from different sources suggests that the
accuracy of the solubility data is not better than 20\%.
The density of vapor was given by the ideal gas equation
of state at 293K: $ \rho=pM/RT=pM/2435 $ where $ p $ is
the pressure and $ M $ the molar mass.}
\vskip 2mm
\hrule
\vskip 0.7mm
\hrule
\end{table}

\begin{table}[htb]

\centerline{\bf  Table 2b\ Vapor pressure and solubility of
water in liquid alkanes (at 20 degree C)
}

\vskip 5mm
\hrule
\vskip 0.5mm
\hrule
\vskip 2mm

\color{black} 

$$ \matrix{
\hbox{Name} \hfill  &
\hbox{Formula} & \hbox{Solubility of water}& \hbox{Vapor pressure} 
 & \hbox{Density of water vapor}\cr
\hbox{} \hfill  &
\hbox{} & \hbox{in alkane}& \hbox{of water in air} 
 & \hbox{in air}\cr
\qtb
\hbox{} \hfill  &
\hbox{} & \hbox{[mg per liter of alkane]}& \hbox{[kPa]}
& \hbox{[mg per liter of air]}\cr
\noalign{\hrule}
\qth
\qtb
\hbox{Water} \hfill  & \hbox{H}_{2}\hbox{O} & 50  & 2.3 & 7.4 \cr
\noalign{\hrule}
} $$
\vskip 0.5mm
Notes: The solubility data of water in liquid alkanes (from pentane
to nonane) are almost independent of the alkane. Surprisingly,
the density of water vapor in air is lower than the density of
dissolved water in alkanes. 
\qL
{\it Sources: Encyclopaedia of hydrocarbons, Section 5.3: Treatment
plants for oil production. p.644. 
Lucia et al. (2012, p. 10).\qL
http://macro.lsu.edu/howto/solvents/pentane.htm (data for pentane at
20 degree)\qL
http://macro.lsu.edu/howto/solvents/hexane.htm (data for hexane at 20 degree)\qL
http://macro.lsu.edu/howto/solvents/heptane.htm (data for heptane at
25 degree)\qL
The density of vapor at a given pressure was given by the ideal gas equation
at 293K: $ \rho=pM/RT=pM/2435 $ where $ p $ is
the pressure and $ M $ the molar mass.}
\vskip 2mm
\hrule
\vskip 0.7mm
\hrule
\end{table}

Fig. 9 refers to the case of hexane but would
be the same for any other liquid alkane.
The data given in table 2a and the graph of Fig. 8 
show that the diffusion has a clear correlation
with the index $ n $ of the alkane. Moreover,
one knows that the attraction between
two alkane molecules increases with $ n $ because
this attraction is due to a (weak) London-type
attraction between the hydrogen atoms.
Thus, the more hydrogen atoms there are, the stronger the
attraction%
\qfoot{More details can be found in Roehner (2004, p. 100-103
and 2005 p. 663).}%
.
In other words, by measuring the diffusion rate
is is possible to estimate the attraction strength, if not
in an absolute way, at least in a relative way provided
the comparison is made between species in which the interaction
mechanism is the same.

%
\begin{figure}[htb]
\ifnum\arxiv=0
\centerline{\psfig{width=14cm,figure=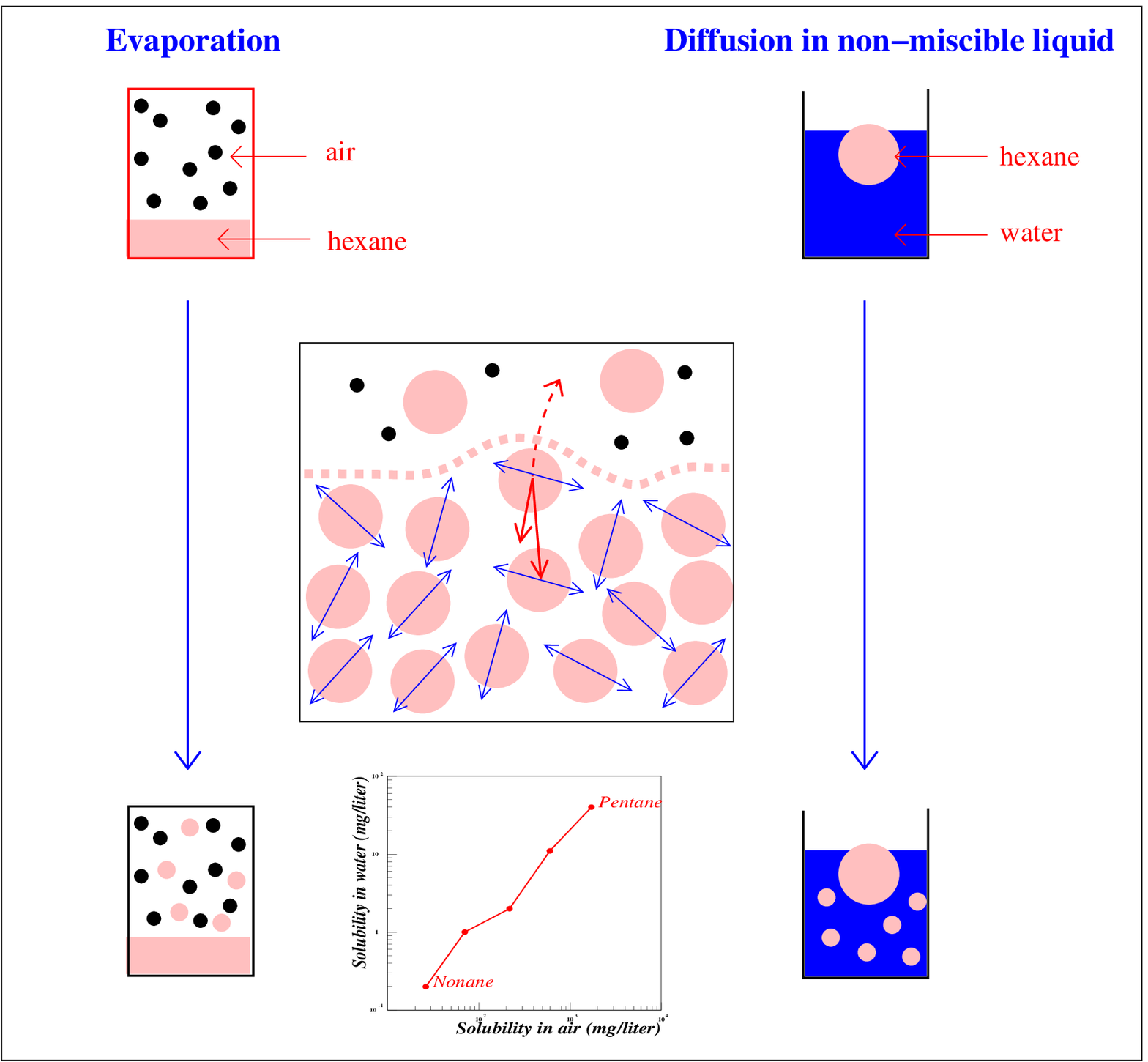}}\fi
\ifnum\arxiv=1 \centerline{\psfig{width=14cm,figure=diffhex.eps}}\fi
\qleg{Fig.\qhu 9\qhv Two similar diffusion mechanisms of a 
liquid $ A $ (hexane) into another fluid $ B $.}
{Left: Evaporation of $ A $; this process 
 can also be called ``solubility in air''.
Right: Diffusion into a liquid $ B $ (water); this process 
can be called solubility of $ A $ in $ B $ (even though the
two liquids are not really miscible).
The two mechanisms have in common the feature
that there is no interaction between the
molecules which diffuse and the particles through which
they diffuse. This similarity is exemplified by the graph
which shows that from pentane to nonane the solubility rates
decrease in the same way. The representation in the middle was
drawn for evaporation but would be very similar for diffusion 
in water. 
The red arrows show the attraction
forces experienced by an escaping molecule that is due to 
the molecules located in the first two layers.
The blue arrows represent the thermal agitation of the molecules.
It is this agitation which allows molecules in the external
layer to escape from the liquid.} 
{}
\end{figure}

\qpar
From the perspective of applying the model of Fig. 9
to living organisms, it can be observed that
whereas the total number of layers
is irrelevant in the case of molecules (because it is very large),
it becomes an important parameter for living
organisms. As long as the distance $ d $ between the external
and deepest layer remains smaller 
than the range of the interaction $ \rho $,
the inward attraction on the external layer
will increase along with the number of layers. This means
that the diffusion rate will become smaller for a larger number
of individuals. On the contrary, once the total number
of individuals is large enough to make $ d $ larger than $ \rho $,
adding more individuals will not change the diffusion rate.
This opens a way for estimating the interaction range,
at least if the background noise is not too large.

\qA{How many ``nearest neighbors?''}

Let us assume that by some method based on diffusion or otherwise
one has been able to estimate the interaction within a system
of euglenas. From these global data one would like to
derive an estimate of the interaction between two euglenas.
How should this be done? 

Once again physics can come to our help.
The conclusion will be that, not surprisingly,  
the answer to this question
completely depends on the number of neighbors with whom
each euglena interacts%
\qfoot{In chemistry, that is to say inside a molecule,
or in crystallography the number of nearest neighbors is
called the {\it coordination number}. For a disordered system
like a liquid, the coordination number cannot be precisely defined.
Instead one will define successive coordination numbers
for spherical shells of increasing radius.}%
. 
This point will become clearer
by examining a specific physical case.
\qpar
Let us consider a liquid, for instance water or ethanol.
A global estimate of its cohesion energy at molecular 
level is given by the heat of vaporization $ \Delta H $.
It is interesting to observe that, according to the so-called
Trouton rule (see the Wikipedia article entitled
``R\`egle de Trouton'' (in French)) $ \Delta H $ is,
with good approximation, proportional to the boiling temperature
expressed in degree Kelvin%
\qfoot{For liquids such as alcohols or water 
which have a H-bond the formula underestimates the vaporization
heat. In those cases it should be replaced by: $ \Delta H =0.11T_b $.
However, if one is only interested in rough and quick estimates
one does not need to take this correction into account.
As a matter of fact, the formula gives acceptable orders of
magnitude even when applied to gases or many solids. 
For instance:\qL
For argon the heat of vaporization is 7.1 kJ/mol whereas
$ T_b=87 $ K gives 7.5.\qL
For aluminum the heat of vaporization is 284 kJ/mol, whereas
$ T_b=2743 $ K gives 238.\qL
But it does not work for sodium chloride whose heat of
vaporization is 790 kJ/mol whereas $ T_b=1686 $ K gives 146.}%
:
 $$ \Delta H (\hbox{\small (expressed in kJ/mole)}=0.087T_b $$

As an illustration we consider hexane whose boiling temperature is
$ T_b=342 $. The previous rule gives $ \Delta H =30 $ kJ/mol, 
which is close to the experimental value of 31 kJ/mol.
\qpar

For water we know both the global and 
molecule-to-molecule cohesion
strength. The first one is given by  $ \Delta H=40 $ kJ/mol
whereas the second one is 16 kJ/mol. The 16 kJ refer
to a mole of pairs of water molecules. The ratio 
$ 40/16=2.5 $ allows us to
answer the question about the number of neighbors. 

%
\begin{figure}[htb]
\ifnum\arxiv=0\centerline{\psfig{width=10cm,figure=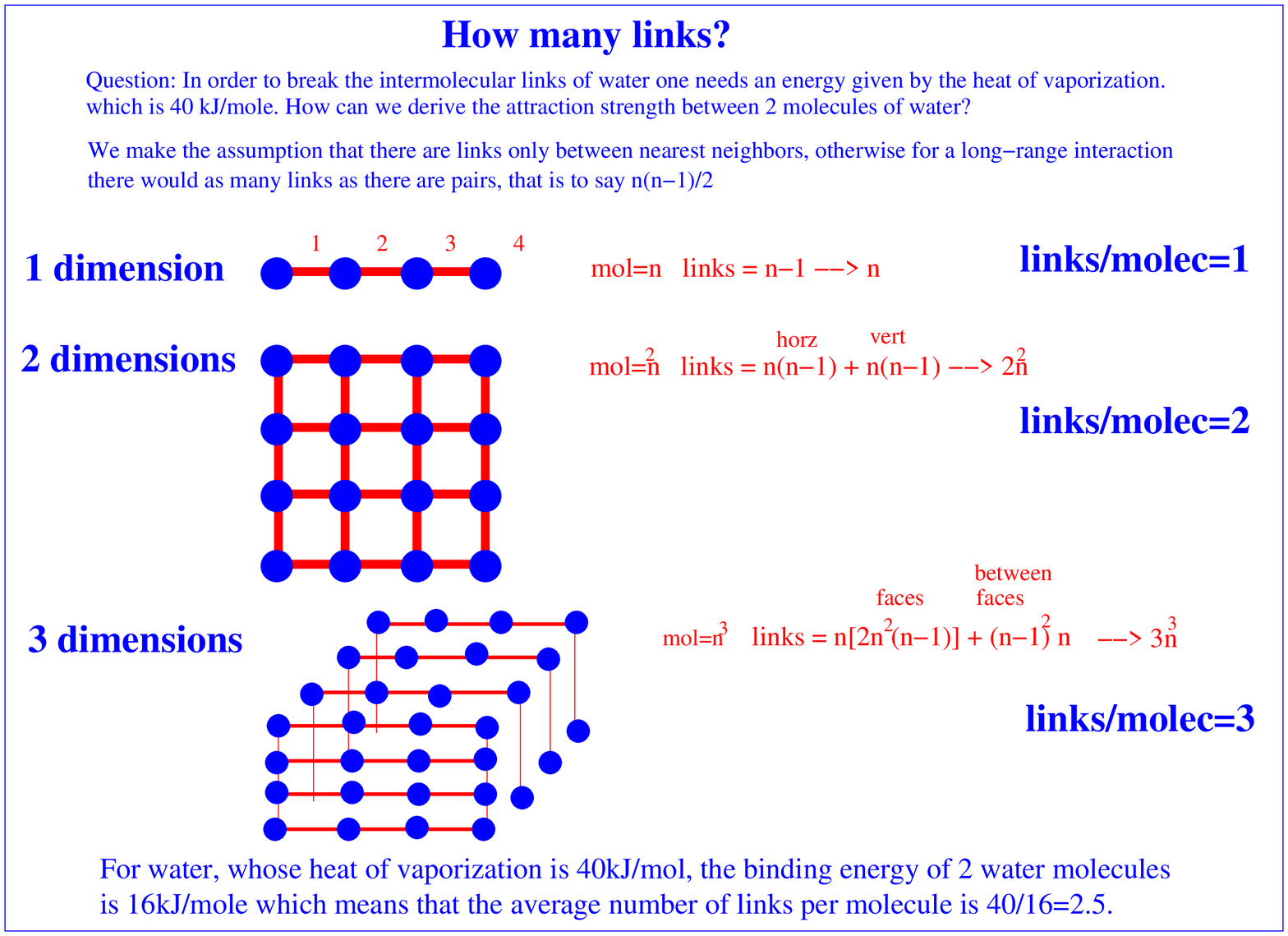}}\fi
\ifnum\arxiv=1\centerline{\psfig{width=10cm,figure=nblinks.eps}}\fi
\qleg{Fig.\qhu 10\qhv How many links per molecule?}
{The number of links per molecule depends upon the
number of spatial dimensions and upon the specific structure of
the system. In order to provide an illustrative case, this 
figure relies on the simplest possible assumption.} 
{}
\end{figure}

How should this result be interpreted? Fig. 10 considers
the fairly ideal case of a rigid simple cubic structure
in which, in addition, there are only interactions between
nearest neighbors. In this special case, each molecule
has 3 links. Naturally, this is a fairly unrealistic case.
In a liquid, the structure is not rigid and the interaction
decreases continuously with distance (as a power $ r^{-6} $ in
the case of van der Waals forces). 
\qpar
In the case of hexane
if we use the same ratio of $ 2.5 $ found earlier for water 
we get: $ 31/2.5=12 $ kJ/mol for the binding energy
between two molecules of liquid hexane.
For heptane, octane and decane one gets similarly:
12.8, 13.6, 15.6. The fact that the interaction increases
with the number of hydrogen atoms comes in confirmation
of an argument made previously.
\qpar
More generally we can write a general formula 
which gives the pair binding energy in any liquid 
(as well as gas or solid) as a function
of its boiling temperature:
$$ \hbox{\small molecule-molecule binding energy (in kJ/mole)}=
3.5\left[{ T_b \over 100 }\right] \quad 
(T_b\hbox{ \small in degree K}) $$ 

The purpose of this formula is to provide orders of magnitude
rather than accurate results.
Its main drawback is that there is
no real way to test the assumption that each molecule
has an average of 2.5 ``nearest neighbors''. 
\qpar
In conclusion, one should favor
cohesion estimates at global level because 
particle-to-particle estimates 
must inevitably rely on shaky structural assumptions. 
\qpar
Sometimes, however, one needs estimates at pair level.
An example is given in the next sub-section.

\qI{Gravity, noise and interaction in diffusion experiments}

In a test-tube we first pour glycerol and then (very slowly to
prevent mixing) water. This experiment raises two questions.
\qbu Will there be diffusion despite
the difference in density (the density of glycerol is 1.26)?
\qbu Will the interaction between glycerol and water molecules
play a role in the speed of diffusion?

\qA{Gravitation versus diffusion}
To answer the first question we must compare the 
gravity potential energy of glycerol and the thermal agitation energy
$ kT $ which brings about diffusion.
In order to avoid small numbers we make the comparison for one
mole which means that $ kT $ becomes $ kN_aT = RT $ where $ R $ 
is the gas constant. \qL
Assuming a height of 10 cm for the liquids, one gets:
{\small
$$ E_p=Mgh=92\times 10^{-3}\times 9.81\times 0.1 \sim 0.1 
\hbox{ J/mol},\quad
RT=8.3\times 293\sim 2.4 \hbox{ kJ/mol } \Rightarrow E_p \ll RT $$
}
Now that we know that there will be a diffusion effect, we can
ask the key-question already examined above: will the diffusion
be interaction-dependent? 

\qA{Diffusion versus attraction}

A quick answer can be obtained
through a ``continuity argument''. For two liquids
whose molecules attract one another strongly (e.g. water-ethanol), 
there will be diffusion. On the contrary, for two liquids
which have no interaction (e.g. water and oil) there will
be no diffusion (except the tiny effect described above).
Therefore, for intermediate cases it is natural to expect
that diffusion will depend upon interaction strength.
\qpar
The question can also be answered in a quantitative way
by comparing $ kT $, the thermal agitation energy of one molecule,
and the interaction energy for a pair of molecules.
The result given above in the previous subsection gives
the interaction energy for water molecules, namely 
$ E_{w-w}=16 $ kJ/mol. We do not know $ E_{g-g} $ or $ E_{w-g} $
but we do not really need to know them precisely. It is 
sufficient to know that they are of the same magnitude as
$ E_{w-w} $. In short: 
$$ E_{\hbox{\small interaction}}\sim { 16/2\over 2.4 }\sim 4\times 
(\hbox{Thermal energy}) $$

In other words, at room temperature 
for a diffusion process involving two liquids 
$ A,B $ 
characterized by interaction energies simular
to those of water%
\qfoot{Because the diffusion is also affected by viscosity
one must also assume that the liquids have comparable viscosity.}%
.
the interaction strength plays a substantial role.
However, whereas
the ratio $ RT/E_p $ was a factor over one thousand, the ratio 
 $ RT/ E_{\hbox{\small interaction}} $ is rather of the order of one.
\qpar

There is a last point that should be added to the many lessons
that physics can teach us. Although perhaps the most important,
it is often overlooked.

\qA{The endless quest for ultimate details}
It is often said that physical phenomena are ``simpler''
than biological or social phenomena. 
\qpar
No matter whether this is true or not, the success of physics
relied on the fact that it first focused on overall understanding,
leaving the intricate details of specific
mechanisms for later investigation.
\qpar
For instance, the detailed mechanisms of inter-molecular 
interactions are not only horrendously complicated
but also of great diversity.
As a matter of fact there are many, many
types of interactions, e.g.
   ion-ion, ion-dipole, H bond, dipole-dipole, ion-induced dipole,
 dipole-induced dipole, London dispersion.
\qpar
Moreover, each one of these interactions has many facets.
Thus, for the H bond one may distinguish the following components:
electrostatic attraction, polarization attraction, covalency attraction,
dispersive attraction, electron repulsion. 
\qpar
Understanding
any of these facets is in itself a formidable challenge both
experimentally and theoretically. In addition, it is not obvious
how scientifically ``rewarding'' 
gaining such an understanding may be.

\qI{Application to the case of living organisms}
Two main  conclusions emerge from the previous analysis.
\qee{1} For strong attraction (ants, bees) there is clustering
instead of diffusion. For low attraction (drosophila) diffusion
leads to uniform density. 
Now, physical diffusion in the low solubility
case suggests that for medium attraction one should
expect a process in which diffusion stops before a uniform
density is reached. Moreover, the 
equilibrium spatial density profile should give a measure of
interaction.  
\qee{2} Until we have got
a good quantitative knowledge of
interaction strength for a broad range of species, one
should not spend too much time on the investigation of detailed
interaction mechanisms. The example of physics shows that
such mechanisms are very complicated and highly diversified. 

\vskip 5mm
\centerline{\LARGE \bf \color{magenta}
Part 4: In search of a general measurement method}
\vskip 3mm

Firstly, we focused on cases in which the attraction
was either very strong (ants,bees, beetles) or very weak
(drosophila). In this way, we were able to propose
methods for estimating attraction strength. However, 
this left open the question for all cases in which
the attraction is not strong enough to produce a
clustering process.
\qpar
In this part we set ourselves a challenge.
We consider a specific species of microorganisms, namely
{\it Euglena gracilis} for which we wish to measure
the interaction strength. As we do not know in advance
whether the interaction is strong, weak or nonexistent
this is a much more difficult task than what we have
done in the first part. It is an ongoing investigation.
We will report it chronologically highlighting
unexpected difficulties. 
We made some progress but
more work is still required in order to reach our 
objectives.

\qI{{\it Euglena gracilis}}

{\it Euglena gracilis} is a unicellular organism 
which swims in water. When moving it
has the shape of a cylinder with
a length of about 50 micrometers and a diameter of about
10 micrometers. Its velocity is
of the order of 1, 2 or even 3 times its length per second.
It is of green color because it contains chlorophyll.
\qpar
Why did we select it? There were practical reasons as well
as scientific reasons.
\qbu The euglenas are easy to keep. A population kept
in a small bottle for over one month will remain
in good shape.
After two months many cells will take on a circular shape and
become fairly static. This indicates that there is need
for a ``revitalizing'' process which requires special expertise,
specific chemical products and techniques. 
\qbu Due to their small size, a drop of euglenas will contain 
thousands of individuals which is a favorable factor
for the investigation of collective phenomenon.
On the other hand, their size is 
large enough to make them observable with a 
stereo-microscope%
\qfoot{Stereo-microscopes always have two eyepieces. Since
stereoscopic vision requires two distinct images, one presented to
the left eye, and one to the right, they must have two objectives.
High power microscopes also have 
two eyepieces but only a {\it single} objective lens.
Because of their small magnification (typically between 20x and 80x) 
stereo-microscopes have a broad observation field which is
an important requirement for our observations.}%
.
\qbu {\it Euglena gracilis} has been studied and used as a model
microorganism at least since 1880. The references of some
early papers and books, e.g.  Engelmann (1879, 1882), Stahl (1880),
Mast (1911), can be found in the reference section. 

\qI{Research plan}

When we started our research in December 2013 our plan was to
proceed in three steps.
\qee{1} First, we wanted to measure the main individual
characteristics, such as sizes, average and standard deviation
of the velocity distribution of the euglenas. 
\qee{2} Secondly, we wanted to see if we could identify expected effects
such as the ``crowded sidewalk'' effect .
\qee{3} Last but not least, we wanted to find a method for
measuring the interaction strength. This is the most important
but also the most challenging part. 
In accordance with the physics-based 
argument given above, our plan was to
try a method based on the diffusion properties of the euglenas.
It is at this point that several surprising observations
made us realize that we had underestimated the sensitivity of
euglenas to low-intensity light. Since, needless to say,
it is impossible to make observation in total
darkness, we realized that it was necessary to study the effect
of light in order to find out what level
of light was acceptable.

Among the surprise-observations that made us change our program,
one can particularly mention the following.
\qee{1}  Euglenas were found to flee light even at low intensity,
e.g. 200 lux, some 10 to 100 times less than what we expected from
the literature.
\qee{2} The diffusion behavior of euglenas is completely changed
even under low intensity light. Basically the diffusion is stopped.
\qee{3} A sudden change in the intensity or color of light 
``freezes'' the euglenas,
a reaction referred to in the literature by the German expression
{\it Schreckbewegung} (shock reaction). 
\qpar
An account of these observations is given below. 
\qpar

Fig. 11 proposes a recapitulation of the range of questions
that arise when one wishes to study the properties
of a system of living organisms.

%
\begin{figure}[htb]
\ifnum\arxiv=0\centerline{\psfig{width=12cm,figure=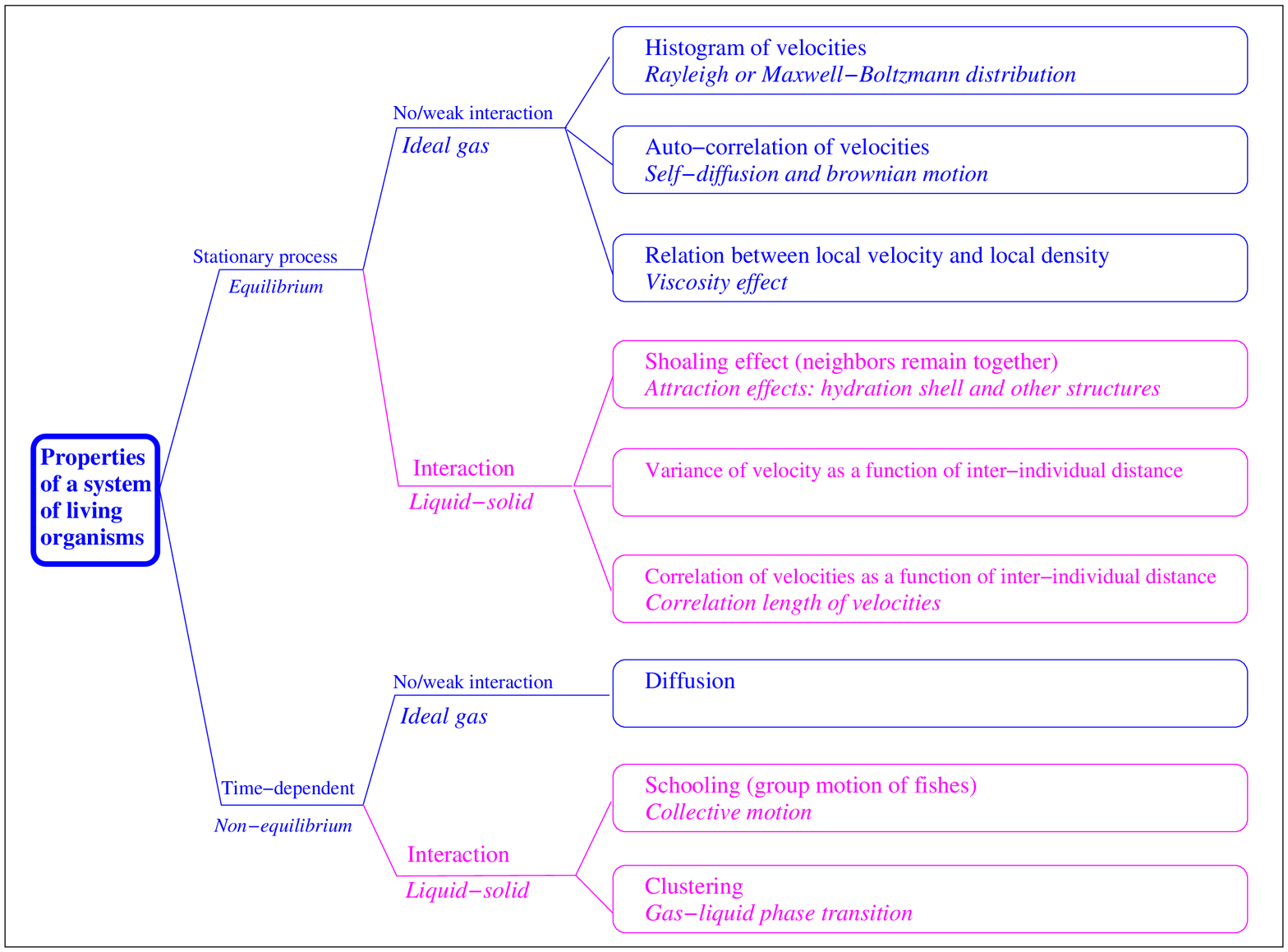}}\fi
\ifnum\arxiv=1\centerline{\psfig{width=12cm,figure=questions.eps}}\fi
\qleg{Fig.\qhu 11\qhv Properties of a system of living
organisms.}
{The questions are arranged in the form of a tree which is
modeled on the classes of questions considered in
statistical physics. These questions are written in italic.}
{}
\end{figure}

The classification takes
its inspiration from physics; it provides a quick
assessment of the expected difficulty of a question
in the sense that:
\qbu  Time-dependent phenomena are more difficult to study than
equilibrium situations.
\qbu Collective properties are more difficult to study than 
individual properties.

\qI{Individual properties}

As an introduction we wish to propose a riddle and to
show a film. 

\qA{Positions and orientations}
Fig. 12 has 4 panels. Two of them have been drawn according
to real pictures of euglenas.  
All little segments have been
drawn in the positions and orientations of the euglenas
on the pictures. On the two other panels the positions and
orientations were selected randomly. For $ x $ and $ y $ the
numbers were drawn from a uniform distribution on 
the interval $ (0,1) $,
and the angular position was drawn from a uniform distribution
on $ (0,2\pi) $. 

%
\begin{figure}[htb]
\ifnum\arxiv=0\centerline{\psfig{width=8cm,figure=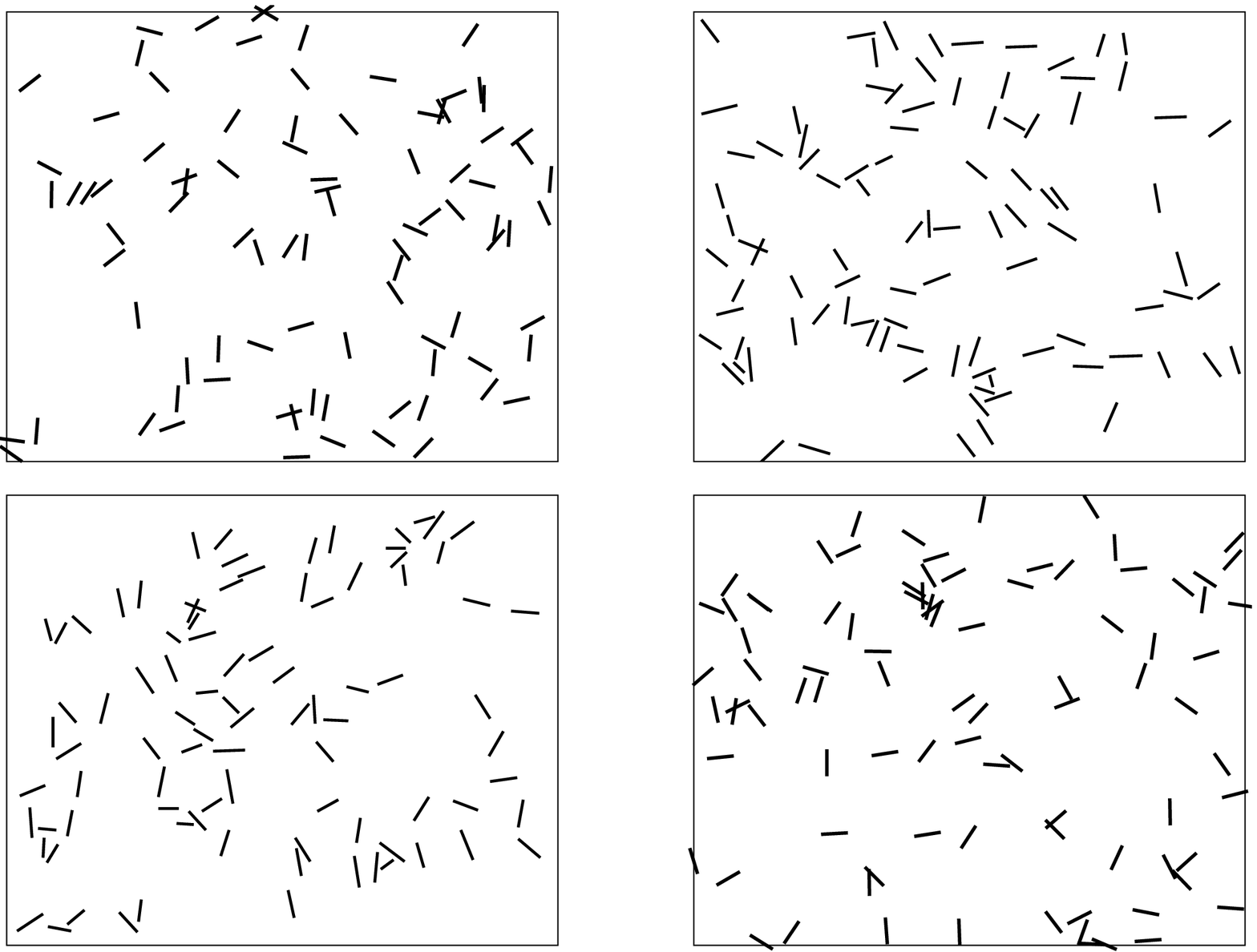}}\fi
\ifnum\arxiv=1\centerline{\psfig{width=8cm,figure=simobs.eps}}\fi
\qleg{Fig.\qhu 12\qhv Simulations versus real pictures.}
{Two of the images are simulations whereas the two other
reproduce the positions and orientations of real
pictures. Is it possible to distinguish between them?}
{}
\end{figure}
Is there a way to distinguish the real pictures from the
simulations? We leave the answer to the reader but
personally we were not able to find a clear criterion%
\qfoot{The pictures of the real euglenas
are 1 (top left) and 4.}%
.
\qpar

Positions and orientations give only a static picture.
By watching a movie%
\qfoot{Available at the following address:
http://www.lpthe.jussieu.fr/$ \sim $roehner/euglena.html}
we will get a dynamic picture. \qL

It reveals some interesting points.
\qbu Euglenas may take two aspects: cylindrical
segment-shaped or disk-shaped. In the second the
euglenas do not move, except perhaps rotating on a vertical
axis.
\qbu Some euglenas move straight ahead. Clearly such deterministic
movements are very different from the erratic zigzags 
of a Brownian particle.
In other words, whereas the positions and orientations of the
euglenas were consistent with a purely random model, their
movements are not.
\qpar

A more quantitative analysis will be to draw the distribution
of velocities.

\qA{Distribution of sizes}
Before that, however,
there is another static property that we need to examine,
namely the distribution of sizes. The distribution
of sizes will affect the distribution
of velocities and if it is fairly broad it will be important
to measure the velocity distribution on samples restricted
to a given size. Fortunately this is not necessary
because the distribution of size is fairly narrow. 
It has a coefficient of variation (i.e. standard deviation/average)
of 0.11 which is much less than the coefficient of variation of
the velocities.

\qA{Distribution of velocities}

The Maxwell-Boltzmann distribution of molecular velocities
in an ideal gas is well known by physicists.
It can be derived theoretically in the framework of the kinetic theory
of gases. Less well known, however,
are the experimental tests of this law. Direct
experimental measurements did not come before the 1930s.
and their accuracy remained hardly better than 5\%. 
\qpar
The pioneer experiment was the observation of the effusion
effect, that is to say the production of a molecular beam,
by Louis Dunoyer (de Segonzac) in 1911. However, its experiment
remained more qualitative than quantitative. The first
experiments that was accurate enough to allow a 
comparison with the M-B distribution
came almost 20  years later first by 
Lammert (1929) followed with a better accuracy by Zartman (1931).
Zartman's results are in good agreement with the
M-B distribution for high velocities but not for the lowest
velocities.  
\qpar
A point of particular interest for the present experiments  
would be to know whether or not the distribution is changed when 
the source is a liquid instead of a gas. A paper by
Johnston et al. (1966) provides a comparison for the case
of helium II but because many corrections had to
be performed the interpretation of the results remains fairly 
unclear.
\qpar

For the experiment with euglenas
we took several movies each of which had 500 images
and corresponded to an observation time of 77s. The
observation field 
measured $ 2.02\hbox{mm}\times 1.52\hbox{mm} $.
The euglenas were moving between microscope slide and 
cover slip and the movie shows that
they could not cross one another. In other words
it was a two-dimensional experiment.
\qpar
For the highest density (18/sq-mm) the observation field
contained about 80 euglenas but some 25\% of them 
were disk-shaped euglenas which did not move%
\qfoot{Only a few
euglenas switched from disk-shape to segment-shape
or vice versa during the 1.54s time intervals used for
measuring the velocities. Therefore this effect was neglected.}%
.   
All the cylinder-shaped euglenas were followed over 10 images
(1.54s). The distance covered was approximated by a straight
line from initial to final point. Then, the same operation was
repeated for a second series of 10 frames 
located about 10 seconds later
so that it may be considered as a realization independent
of the first one. This procedure was iterated several times. 
\qpar

%
\begin{figure}[htb]
\ifnum\arxiv=0\centerline{\psfig{width=15cm,figure=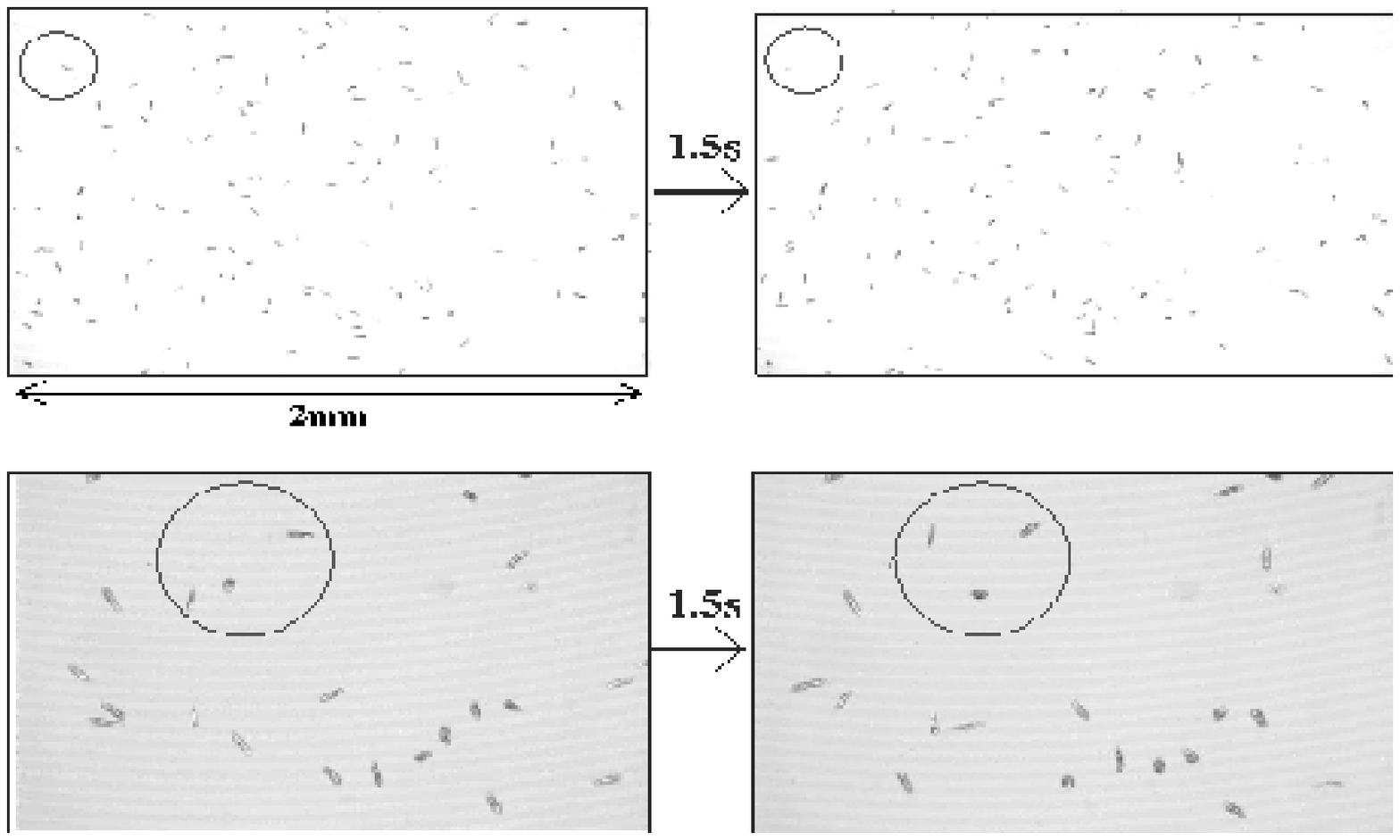}}\fi
\ifnum\arxiv=1\centerline{\psfig{width=15cm,figure=eugv.eps}}\fi
\qleg{Fig.\qhu 13a\qhv Pictures of euglenas over a time interval
of 1.5 second.}
{The two pictures in the first line show the sort of
pictures used for measuring the distribution of velocities. 
The red circles identify the movement of one specific euglena.
The pictures in the second line provide a better view of the
shape of the euglenas but for this higher magnification the number
of euglenas contained in the observation field is too
small for the purpose of measuring velocity distributions.}
{Source: The observations were performed in December 2013 
in the ``Cell cycle and cell determination'' group 
of University Pierre and Marie Curie.}
\end{figure}
Prior to analyzing the distribution of velocities,
there was a first useful check to be made which was the following.
Needless to say, to take the movie the sample had to receive
low intensity of light coming from below the sample. 
In order to check whether or not the light had an influence on the
motions of the euglenas (for instance due to non-uniformity)
we computed the sum of the velocities of all euglenas on
a given frame. 
A non-zero sum would reveal a drift due to
an asymmetry of the light-source. In fact, the averages of the
$ x $- and $ y $-components were less than 5\% of the average
distance covered over the 10 frames. Nevertheless a correction
was performed by subtracting the drift before drawing the
velocity distribution.   
\qpar

Before studying the modulus of the velocities, we studied
their $ x $- and $ y $-components. Their distributions
turned out to be very well approximated by a Gaussian
distribution.
The test was performed through the so-called
``Henry method'' (see the Wikipedia article in French
entitled ``Droite de Henry''). It is based on the fact that
the values of the 
inverse function of the cumulated normal distribution
as a function of the values of the random variable should
be a straight line if the distribution is Gaussian.
It was indeed a straight line with a coefficient of linear
correlation equal to 0.995. What will be the implication
of this result for the distribution of the modulus of the
velocity? 
\qpar
In probability theory there is a theorem (see for
instance Parzen 1960, p. 324) which says that 
if $ X_1, X_2, \ldots ,X_n $ are independent random variables
each normally distributed with mean 0 and same variance, then
$ Y=\sqrt{X_1^2+X_2^2+\ldots + X_n^2} $ follows a 
$ \chi $ distribution. \qL
Moreover, from the properties
of the $ \chi $ distribution%
\qfoot{See for instance the Wikipedia article entitled
``Chi distribution''}
one knows that its coefficient of variation is given
by the following formulas%
\qfoot{The fact that the coefficient of variation does 
does not depend upon the variance of the Gaussian variables
is not really surprising. It is due to the positivity of $ Y $;
as a result, an increase in the variance will also
increase the average. Moreover, it can also be shown  
that when $ n\rightarrow \infty $,
the coefficient of variation converges toward 0:
$ \hbox{CV}\sim 0.7/\sqrt{n} $.}%
:
{\small
$$ 
CV_{n=1}=\sqrt{\pi/2-1}\simeq 0.75,\quad 
CV_{n=2}=\sqrt{4/\pi-1}\simeq 0.52,\quad
CV_{n=3}=\sqrt{3\pi/8-1}\simeq 0.42
$$
}
\qbu The case $ n=1 $ would correspond to euglenas moving in a
narrow capillarity tube with a diameter the size of the
euglenas. 
\qbu The case $ n=2 $ corresponds to euglenas  
moving between slide and cover slip as in our observation.
In this special case the $ \chi $ distribution is called
a Rayleigh distribution.
\qbu The case $ n=3 $ corresponds to
3-dimensional movements; it can be either euglenas or 
the molecules of a gas. In this special case
the $ \chi $ distribution is called a Maxwell-Boltzmann
distribution.   
\qpar

%
\begin{figure}[htb]
\ifnum\arxiv=0\centerline{\psfig{width=15cm,figure=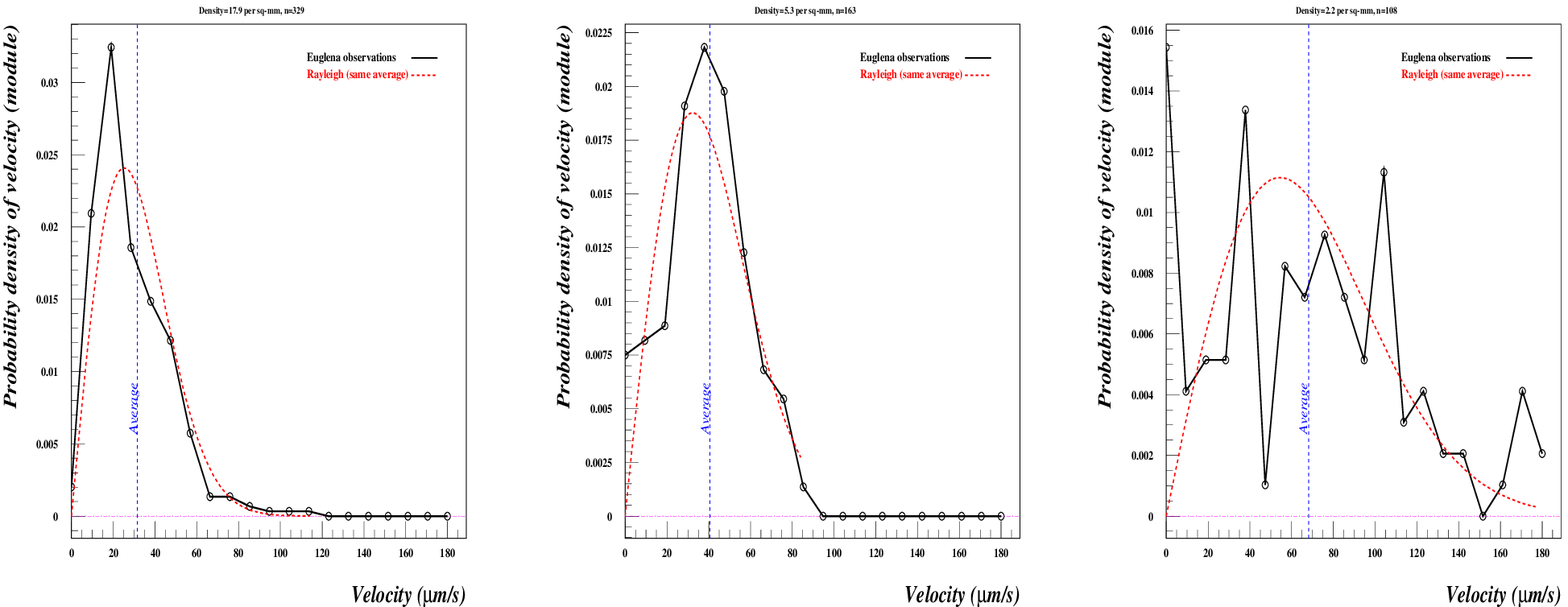}}\fi
\ifnum\arxiv=1\centerline{\psfig{width=15cm,figure=dv.eps}}\fi
\qleg{Fig.\qhu 13b\qhv Velocity density distribution of euglenas.}
{The 3 graphs from left to right correspond to decreasing
densities: 18/sq-mm, 5.3/sq-mm, 2.2/sq-mm. To these decreasing
densities correspond increasing average velocities: 31 micro-m/s,
40 micro-m/s, 68 micro-m/s. The coefficient of variation does
not show any clear trend. Its average for the 3 cases is 0.55,
not far from the coefficient of variation of a Rayleigh distribution
which is $ CV=\sqrt{4/\pi-1}=0.52 $.}
{Source: The observations were performed in December 2013 
in the ``Cell cycle and cell determination'' group 
of University Pierre and Marie Curie.}
\end{figure}
 
From the results shown in Fig. 13 it appears
that the coefficient of variation is not far from the
value 0.52 corresponding to the Rayleigh distribution.
One can also see that, in accordance with what is
expected from the crowded sidewalk effect,
the mean speed decreases when the density increases .

\qA{Velocity distribution for ants}
A similar study based on data for fire ants published in a book
by Pr. Deborah Gordon (2010, fig. 3.1) 
led to a coefficient of variation of 0.55.

\qA{Temperature effects on movement}
It is well known that the average speed $ v_m $ of the molecules of a
gas increases with Kelvin temperature $ T $: $ v_m\sim \sqrt{T} $.
It would not be surprising to observe a similar effect for
living organisms. For instance it is well known that the activity
of ants or bees is fairly low when the temperature is below 10 degree
Celsius. Such an effect has also been observed for
groups of small fishes by students at Beijing Normal University.
\qpar

A similar study for bacteria can be found in Schneider and Doetsch
(1977, p. 697). They studied 4 species of bacteria%
\qfoot{{\it P. mirabilis, S. typhimurium, S. serpens, P. fluorescens}}
over a temperature range from 10 to 40 degree Celsius.
For each species,
the average speed increased with the Celsius 
temperature $ t $ like a function
of the form $ t^{\alpha},\ \alpha>1 $. Over the whole
range the velocity (averaged over the 4 species) was multiplied
by $ 6.1 $. Such an increase is of course much faster than
the increase of $ \sqrt{T} $.

\qI{Relationship between local density and speed}

In the previous section we have seen that for {\it whole
samples} of different densities the average speed decreases
when the density increases. In Fig. 12 we have also seen that
within a given sample the density is not uniform.
This leads us to the question of whether it is possible to
identify a connection between local density changes and
changes in spatial average of velocity.

\qA{Methodology}
In its principle
the methodology is straightforward. We divide each image
into a number ($ p $) of zones, then we compute
the averages of both the density ($ d_m $ )
and the velocity ($ v_m $) in each zone; finally, we test
the correlation of the $ (d_m,v_m) $ points. The main difficulty
is to implement this procedure 
in a way which minimizes the background noise.
\qpar  
More specifically, we divided each image into $ 3\times 2 =6 $
zones%
\qfoot{The images have 696 pixels in the $ x $ direction and
520 in the $ y $ direction.}%
.
Why 6 zones? \qL
It appears to be the best compromise between
two conditions which can be explained as follows.
In order to minimize the fluctuations of the averages,
one needs as many euglenas as possible in each zone.
This implies big zones. On the other hand however, each
zone (over the time interval required to compute the velocities)
will give only one $ (d_m,v_m) $ data point. Thus, big zones
will give few data points and in addition
they will have a fairly narrow dispersion. 
This, in turn, will lead to
large confidence intervals for
the correlations. On the contrary, small zones will give a larger
number of fairly ``noisy'' data points.
In order to find the best compromise we tried several subdivisions.
The division into 6 zones turned out to be the best choice.
With this choice there were about 10 euglenas in each zone.

\qA{Results}
Altogether, based on 62 $ (d_m,v_m) $ data points, one gets
the following correlation between density and velocity:
{\small
$$ \hbox{density-velocity correlation:}\quad
r=-0.55,\quad \hbox{confidence interval (0.95 confidence level)}
=(-0.7,-0.2) $$  
}
As zero is outside the confidence interval the correlation
can be said to be significant.

\qA{Is the (density,velocity) connection a self-reinforcing mechanism?}
The previous correlation implies that
when an euglena arrives in a high density zone it will
slow down. As a result, by spending in this zone more time
than elsewhere, it will contribute to the crowding.
If this is really a self-reinforcing mechanism it should result
in spatial density disparities that are sharper than those
which result from the mere effect of randomness. 
\qpar

In order to probe this effect one would need a series of pictures
(taken almost at the same moment) covering a large area 
of the sample. Why can we {\it not} use frames (sufficiently
apart in time) of a film focused just on one narrow area?
The reason can be understood by taking a look at the 
figures given above for the clustering process for ants. 
In this case  
the self-reinforcing mechanism is made obvious when we watch
the whole population because we see a number of clustering centers.
However, if one restricts the observation to a narrow area far
from the clustering centers it is likely that the density
pattern will not be very different from one based on purely
random numbers.      
\qpar

The observation of a broad spatial area will be one of our objectives
in forthcoming experiments. To begin with, this can be done
in a capillarity tube. However, one needs to be very careful
about possible disruptive effects of light. The perturbations
brought about by light are described in the next section.

\qI{Reactions brought about by light}

We describe 4 experiments which show various reactions 
of the euglenas to light or lack of it. 

\qA{Positive versus negative phototaxis}
Phototaxis is the ability to move in response to light.
The broad rule is that euglenas are attracted by weak light
and on the contrary repulsed by strong light. However, it seems
that there is no clear agreement in the existing literature
regarding the threshold between the two types of behavior.
This may be due to the fact that the reaction of the euglenas
to some extent depends upon their previous exposure.
Although the objective of the experiment described in Fig. 14
was more qualitative than quantitative it sets the
boundary between the two responses somewhere around 100 lux%
\qfoot{1 lux corresponds to a full moon night,
150 lux is what is needed to read, 130,000 lux corresponds
to sunlight on a bright summer day. The last case
corresponds to 1000 W/sq.m which gives an equivalence
between the two units valid for sunlight.}%
.

\begin{figure}[htb]
\ifnum\arxiv=0\centerline{\psfig{width=15cm,figure=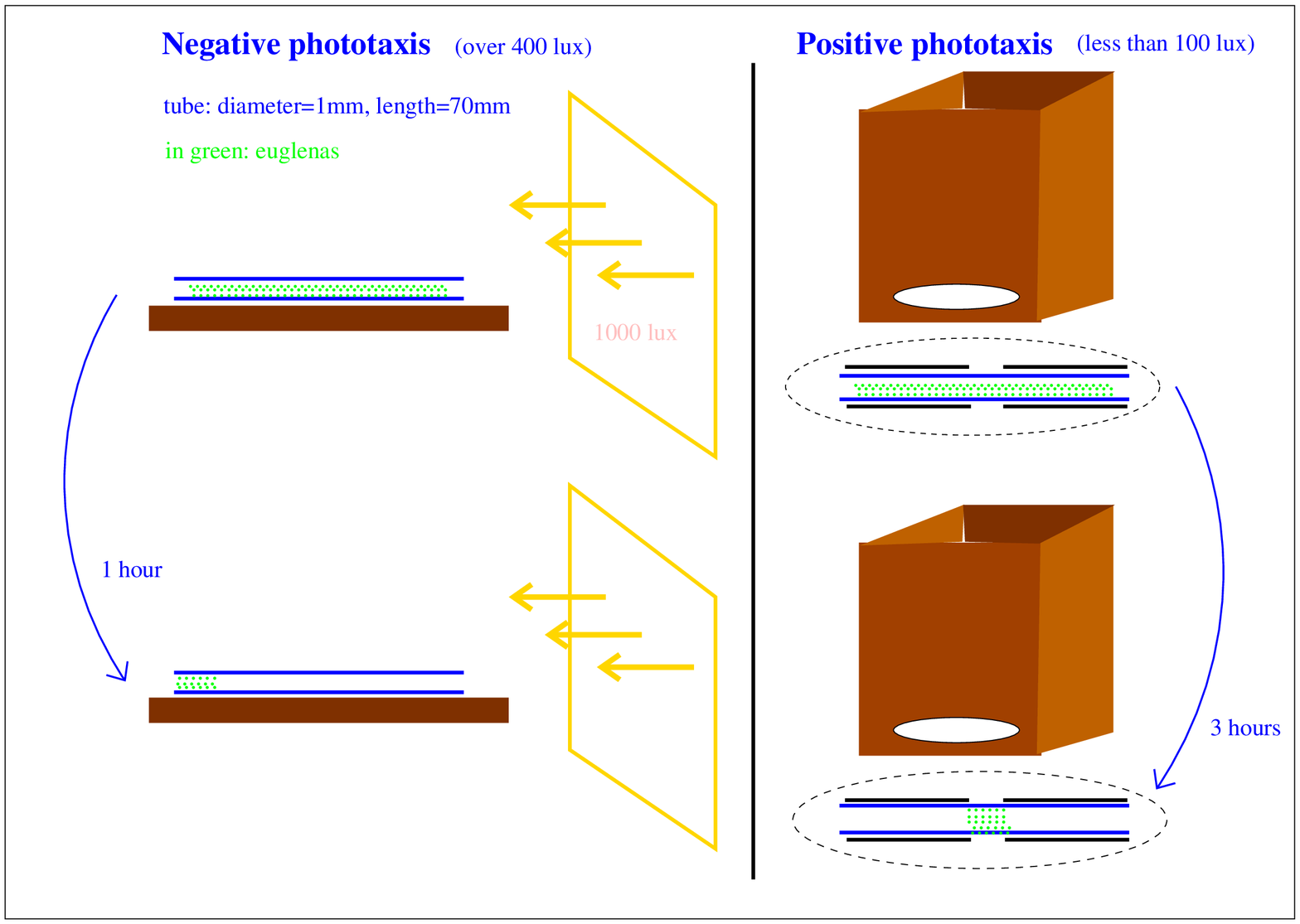}}\fi
\ifnum\arxiv=1\centerline{\psfig{width=15cm,figure=euglight1.eps}}\fi
\qleg{Fig.\qhu 14\qhv Positive versus negative phototaxis.}
{The tube containing the euglenas have an internal diameter
of about 1mm. In the  experiment on the right-hand side
a sheath recovered the tube except for a 5mm interval
around the middle of the tube which could receive
the faint light that could reach the bottom
of the box.}
{Source: The observations were performed in March-April 2014 
at the LPTHE (University Pierre and Marie Curie).}
\end{figure}
 
It can be noticed that the gathering of the euglenas in the
positive phototaxis experiment was about three or four times
slower than in the negative phototaxis experiment.

\qA{No diffusion in dim light versus diffusion in darkness}
When a batch of euglenas is introduced into a capillarity
glass tube (1.15mm in diameter) their behavior is very different
depending on whether the tube is put under uniform dim light
or in complete darkness. In the first case, there is almost
no diffusion, whereas in darkness the euglenas spread
to the whole tube within 30mn approximately.
\qpar
This is illustrated in Fig. 15. The first three panels concern
the case with light. In the last panel the green curves
show the frozen diffusion under dim light whereas the black
curves show fast diffusion in darkness. 
\qpar

%
\begin{figure}[htb]
\ifnum\arxiv=0\centerline{\psfig{width=15cm,figure=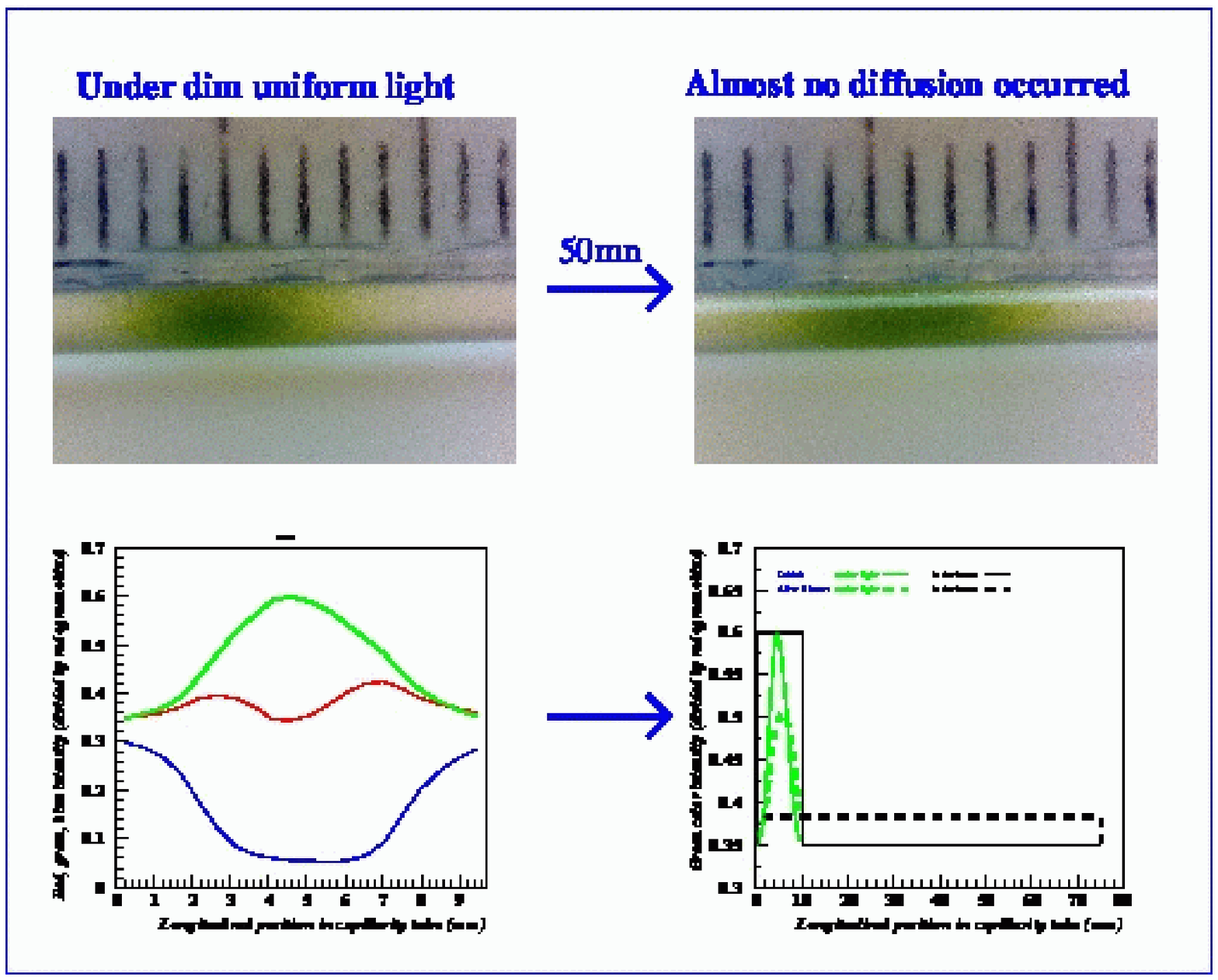}}\fi
\ifnum\arxiv=1\centerline{\psfig{width=15cm,figure=eugdif.eps}}\fi
\qleg{Fig.\qhu 15\qhv Negligible diffusion under dim light
versus quick diffusion in darkness.}
{The two top panels show that under dim light
there was almost no change in the
concentration of the euglenas. A quantitative confirmation
was obtained by analyzing the longitudinal distribution
of the green color using software commands from
ImageMagick. Panel 3 shows the spatial distribution of red, green and
blue light at the start of the experiment. The last panel shows
that after 50mn
the concentration profile (measured by the green color)
is almost the same as at the beginning. 
The internal diameter of the capillarity tube was 1.15mm
and its length was 7cm. The light provided by the
microscope had a fairly uniform intensity throughout the tube
length.\qL
On the contrary, in complete darkness there was a marked diffusion
process with the result that after about 30mn the euglenas
occupied the whole length of the tube. This is represented by the black
curves: the solid line is for the beginning of the experiment
whereas the low dotted line represents the fairly uniform distribution
after 30mn. These curves are rather schematic because we avoided
light as much as possible.}
{Source: The observations were performed on 31 March and 1 April 2014 
at University Pierre and Marie Curie.}
\end{figure}

So far we do not have any explanation for 
the experiment described here. It can hardly be explained 
in term of phototaxis because it is characterized by a lack
of motion. 
\qpar
Another behavior described by early observers
(Engelmann 1882, Mast 1911) consists in the fact that
a light spot acts as a trap for the euglenas: once they
have entered it they cannot leave it (like a night
butterfly which flies around a lamp). 
However this does not apply here for there was  
fairly uniform, low intensity
light on the entire capillarity tube. 
\qpar

One question remains unclear: was the light level really
uniform? In other words, was the luminance the same in all
directions? In appearance it seemed so. However, the fact
that we used two fibre-optic light sources, one on each side,
raises some questions.
Although these sources were at least 10cm distant from the
sample it is possible that there were two
opposite light gradients. If this is the right
explanation for the frozen diffusion, and
as these gradients were certainly fairly
small, it shows that the euglenas are sensitive even to
slight light gradients.\qL
In the future the experiment
will be repeated in different conditions. 

\qA{Effect of a sudden change in light intensity: 
{\it Schreckbewegung}}

When the light is suddenly changed from red to blue the euglenas
stop moving, take on a circular shape (perhaps trying to
minimize their exposed area). This occurs very quickly in less than
0.3 s. As this behavior was first discovered and described by
German researchers it became known as {\it Schreckbewegung} which 
means shock reaction. It can be noted that as the energy of
red light is about one half of the energy of blue light%
\qfoot{$ E=h/\lambda $, where $ h $ is the Heisenberg constant
and $ \lambda $ the light wavelength.}
a switch from blue to red amounts to a reduction in light
intensity. Moreover the euglenas may also be more sensible to
some wavelength intervals than to others.

\qA{Light induced clustering}

Fig. 16 describes two experiments in which the clustering
of the euglenas is brought about by their exposure to strong light.
What are the similarities and differences between the two experiments?
\qbu In both experiments the light comes from below.
Another similarity is the reaction time which is of the
order of 5mn.
\qbu In the second experiment the height of the container
is about 20 times smaller than in the first.  
\qbu Another difference is that in the second experiment
the light is applied only to a fraction of the container area,
the rest of it receiving whatever light comes
(through diffusion and reflection) from the lighted zone.
\qpar
  
When the light is turned on
the euglenas in the lighted zone have a reaction
similar to the {\it Schreckbewegung} described previously.
Yet, the cluster does not appear in this zone
but in its darker neighborhood.   
The experiment was repeated several times with same results.

%
\begin{figure}[htb]
\vskip 3mm
\ifnum\arxiv=0\centerline{\psfig{width=15cm,figure=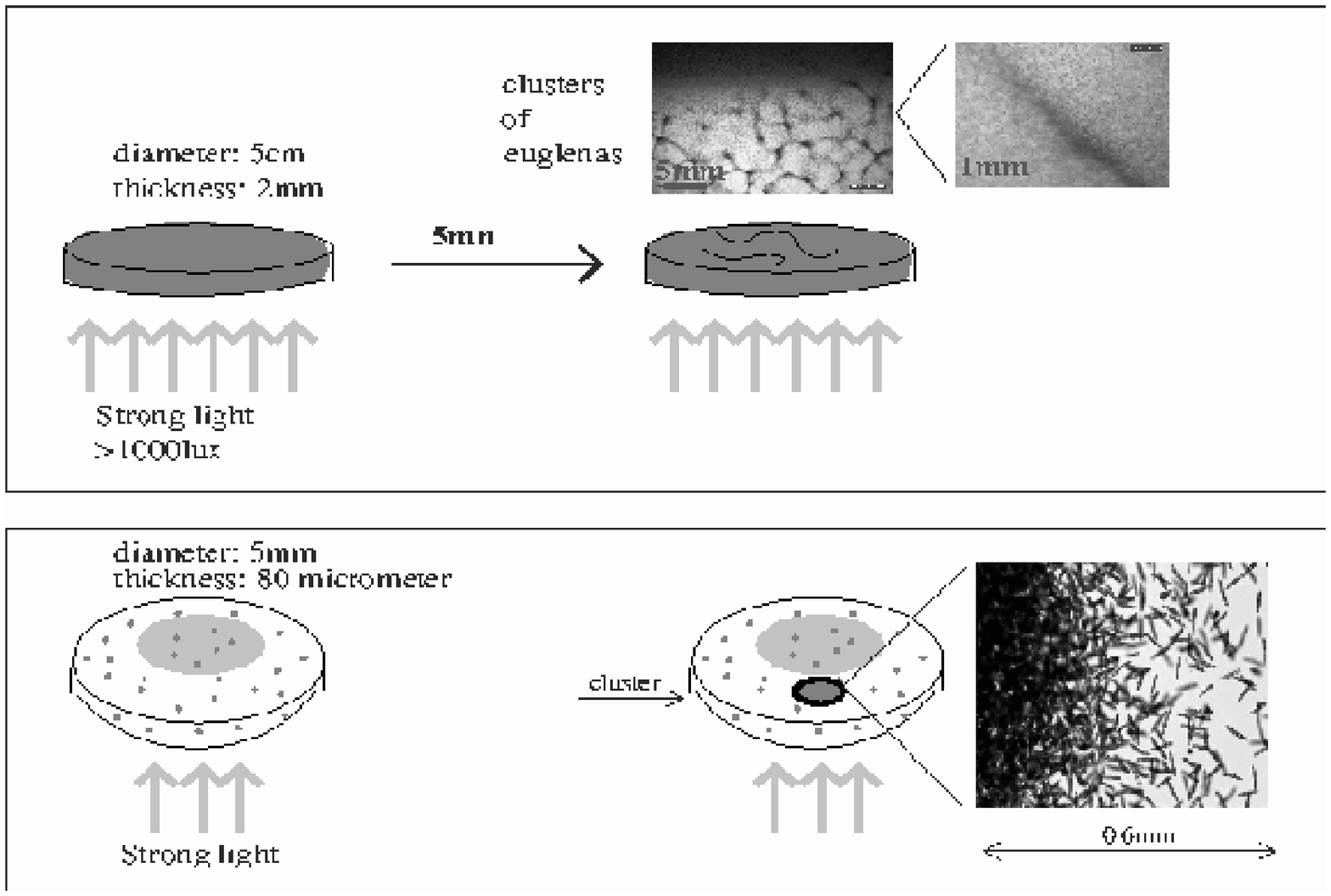}}\fi
\ifnum\arxiv=1\centerline{\psfig{width=15cm,figure=euglight2.eps}}\fi
\qleg{Fig.\qhu 16\qhv Clustering brought about by exposure
to strong light.}
{In both cases the clustering process takes about 5mn 
which means that it
is some ten times faster than the clustering of ants
described previously.}
{Source: The first observation was analyzed in detail in
Suematsu (2011). The pictures were taken in December 2013
and March 2014 at Beijing Normal University
and University Pierre and Marie Curie respectively.}
\end{figure}

\qI{Evidence of interaction}

There are two fairly different methods for the identification
of interactions between the euglenas.
\qbu The first method is to analyze a series of pictures taken
at sufficiently short time intervals to allow a measurement
of instantaneous velocities. The hope is that the movements of the
euglenas will reveal some kinds of pattern which are typical
of interactive individuals, For instance, if the correlation between
their velocities  decreases for larger separation between them
one might be think that it is a result of their interaction.
However, it can also be a consequence of the side-walk effect. 
Indeed,  if 
the euglenes slow down in areas of higher
density this will result in a separation-dependent correlation.
\qbu The second method is to observe a form of collective motion that
cannot easily be explained without interaction. 
\qpar
We have tried the two methods. In the films in which the
euglenas were between slide and cover slip we never saw any clear. 
pattern of collective behavior. It was probably due to the fact
that there was only one layer of euglenas. Another factor is the fact
that when working with a substantial magnification the 
observation field is so small that one may well miss an existing collective
pattern.\qL
One of the clearest results
obtained in analyzing the
movements was the relationship between average velocity and
local population density. The other tests were rather inconclusive.
\qpar

In the two following subsections we describe two collective movements:
(i) the formation of clusters (ii) the formation of networks.

\qA{Aggregation of euglenas}
As already said, the formation of high density areas can be explained
by the side-walk effect. However, the clusters shown in the 
pictures are extreme forms of high density areas in the sense 
that the euglenas form a solid bundle in which they seem to be
completely steady; in fact the bundle is so compact that it is
difficult to see what happens inside. The main point is that
without any attraction, the euglenas would just swim around
the clusters whereas on the contrary one sees a permanent flow of
euglenas entering and leaving the cluster.. 
In other words, one has the
feeling that such clusters cannot be explained through the side-walk
effect.
\qpar
%
\begin{figure}[htb]
\vskip 3mm
\ifnum\arxiv=0\centerline{\psfig{width=15cm,figure=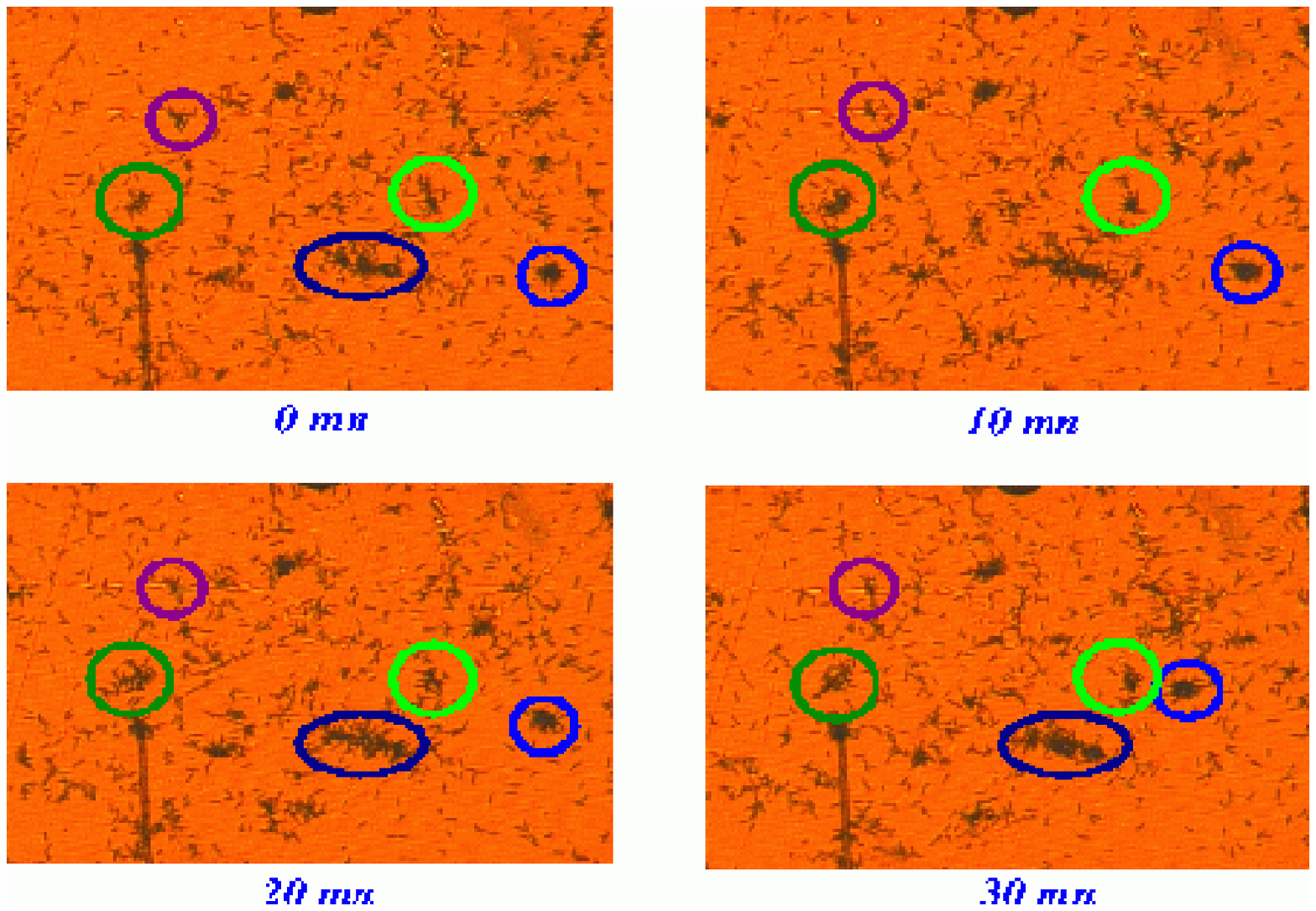}}\fi
\ifnum\arxiv=1\centerline{\psfig{width=15cm,figure=agreg2.eps}}\fi
\qleg{Fig.\qhu 17\qhv Small- and medium-sized clusters of euglenas.}
{When euglenas are observed between slide and cover slip
there is only enough space for
one layer of a thickness around 70 micro-meters; in such conditions
they cannot cross one another and (for some unknown reason)
do not form any stable clusters. In the present experiment 
the euglenas were {\it not} between slide and slip. 
They were in a kind of  mini-swimming pool which was
2mm wide, 100mm long and 0.4mm deep. The pool 
had been made watertight to prevent evaporation.
For most of the time the euglenas received
very little light (less than 30 lux) except during the short
time intervals when pictures were taken. 
During these moments the light 
came from above for it was an inverted microcope that was
used.
The circles and ellipses show clusters of various sizes.
The cluster inside the light blue circle can be seen to move around
globally. Also of interest
is the fact that the mini-cluster inside of the magenta circle
almost disappeared.}
{Source: The experiment was done on 14 May 2014 at the University
Pierre and Marie Curie (IMPMC)}
\end{figure}

There are two other mechanisms that are mentioned in the literature
for explaining the formation of clusters: one is self-shading
and the other bioconvection. \qpar
The conditions of the present experiment were chosen
with the purpose of eliminating these effects. \qL
To eliminate self-shading the
euglenas were left all the time in low light (less than 30 lux)
except when
the pictures were taken. \qL
To eliminate bioconvection that is to say
vertical up and down movements of the euglenas we put them
in a swimming pool whose depth was only 0.5mm. This would
allow several layers of euglenas but would
likely be too narrow to have any substantial vertical movements.
In addition, whereas bioconvection experiments involve a bright
light source beneath the euglenas, 
in our experiment there was none.

\qI{Network formation}

\qA{Observation}
Observation shows show that, in some conditions,
the euglenas form a kind of lattice.

\begin{figure}[htb]
\ifnum\arxiv=0\centerline{\psfig{width=15cm,figure=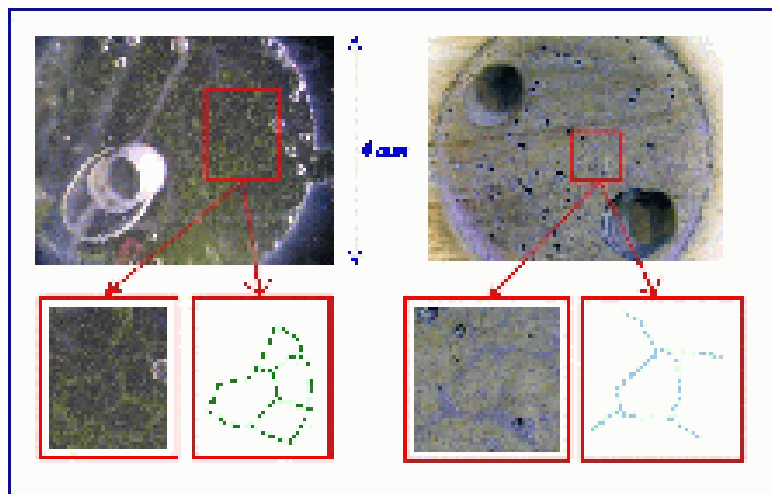}}\fi
\ifnum\arxiv=1\centerline{\psfig{width=15cm,figure=reseau.eps}}\fi
\qleg{Fig.\qhu 18a\qhv Formation of networks of euglenas.}
{The ``swimming pool'' of the euglenas comprised two
circle-areas connected by a narrow canal. However, in the 
present formation of networks the communication
canal probably did not play any role for
each circle-area seemed to change independently from the other.
The depth of the pool was 0.5mm which means that there were certainly
several layers of euglenas. The network appeared some 20mn after
the device had been filled (at 11:15 am). It subsisted almost unchanged for
several hours but disappeared during the night.
At 9:00 am on the following morning
all euglenas were on one side
in the left-hand side circle-area. In this area
their density was fairly uniform.
The image on the right-hand side is a negative version of the original
picture; moving over to the negative slightly improves
the contrast. The origin of the small bubbles is unclear. They
disappeared a few hours later. The 3 big circles 
(one on the left and two on the right-hand side) are the
holes (of diameter 1mm)
through which the euglenas were injected.\qL
The same observation can be made when the depth of the
swimming pool is 0.1mm that is to say about twice the length
of the euglenas.}
{Source: The experiment was done on 2 June 2014 at the University
Pierre and Marie Curie (IMPMC). Many thanks to Dr. C\'eline
F\'erard for her help.}
\end{figure}

 Can the formation of
such a lattice be interpreted as a proof of the existence
of inter-individual interactions? The best we can do to
get a clearer insight is to discuss briefly other
(better known) cases of lattice formation. We will
start with the simplest case and then move by steps
to the more complicated cases.
\qpar
First we will
discuss the static situation of an isolated system
which occurs in the formation of crystals. Then we discuss
the formation of snowflakes which is 
a static situation for a system that is not isolated.
Finally we discuss bioconvection which is a dynamic
effect in a system subject to an exogenous factor (light). 
\qpar

\qA{Crystals}
In physics, more precisely in crystallography, 
observed lattice patterns are the result of
inter-atomic interactions. 
\qpar
As an illustration, one can mention
the case of solid sodium chloride, NaCl, which forms a simple
cubic lattice in which sodium and chloride ions alternate
with each other. Every 
positive sodium ion $ \hbox{Na}^{+} $ 
is surrounded by 6 negative chloride ions $ \hbox{Cl}^{-} $
and vice versa.  
The surrounding
ions are located at the vertices of a regular octahedron.
This lattice is called a face-centered cubic lattice.\qL
There is of course 
an electrostatic attraction between the positive and
negative ions. However, the fact that a $ \hbox{Na}^{+} $ ion
is in equilibrium (except for small vibrations) should not
be interpreted as resulting from the fact that the
attractions of the 6 chloride ions cancel each other
for this would lead to an unstable equilibrium.
In fact, the equilibrium results from the existence, in
addition to the electrostatic interaction, of 
a strong short-range repulsive force. \qL
One may wonder why sodium chloride forms a face-centered lattice
whereas cesium chloride forms a body-centered lattice.
The reason is in relation with the size of the ions and 
the range of their ionic interaction. The space allowed
for the sodium ion is determined by the ion radius of the
sodium chloride and by the geometry of the lattice.
Sodium has 11 electrons, whereas cesium has 55.
Thus, to fit into the crystal, cesium needs a different geometrical
arrangement than sodium.
\qpar

In short, one can keep in mind that for an isolated
system the formation of a lattice is due
to inter-individual interactions and that the shape of the
lattice reflects individual characteristics such as
size and interaction range.

\qA{Snowflakes}

For a system subject to exogenous factors the situation
is more complicated. An obvious illustration is given
by the shape of snowflakes. Although all snowflakes 
consist of crystals of ice, they occur in a great
variety of shapes. Why?\qL
Several exogenous factors such as temperature or the
density of water dropplets will affect the formation
of snowflakes. Moreover, between their formation and the moment
when they become big enough to be
released from the cloud, snowflakes experience a growth process 
which can last from a few minutes to several hours depending
on conditions%
\qfoot{In laboratory conditions 
for a tiny crystal of ice to grow into a snowflake
may typically take about 45mn.}%
.
The final shape will reflect both the shape of the seed-crystal
and the conditions experienced during the duration of growth.
For instance, initial crystals with a thin edge grow faster,
ultimately leading to large, thin flakes, whereas
snowflakes that begin with blunt edges grow more slowly and 
eventualy lead to small, thick flakes.

\qA{Bioconvection}

Ever since the first systematic observations done by Wager in 1911,
the lattices formed by the euglenas were attributed to
the phenomenon of bioconvection. The models of bioconvection
rely on the so-called Rayleigh-Taylor instability.
This hydrodynamic effect occurs when a liquid of density $ \rho_1 $
forms an upper layer atop of a liquid of lower density 
$ \rho_2 <\rho_1 $.
For instance, in the experiment described in Plesset and Winet (1974)
$ rho_1=1.008 $ (corresponding to a concentration of euglenas 
of $ c_1=1,400 $ per cubic-mm) and $ rho_2=1.007 $ (corresponding
to $ c_2=560 $ euglenas/cubic-mm). 

\begin{figure}[htb]
\ifnum\arxiv=0\centerline{\psfig{width=12cm,figure=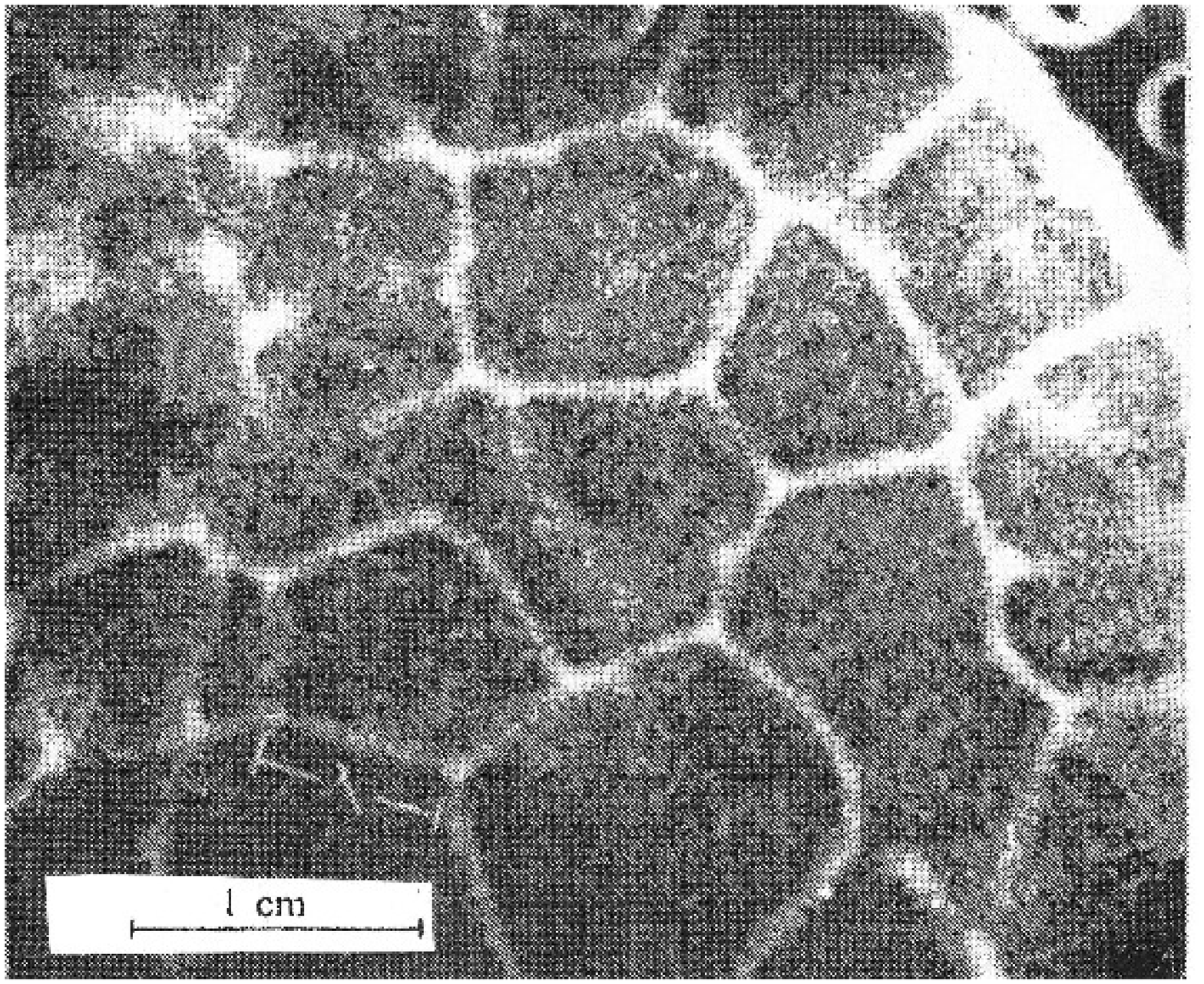}}\fi
\ifnum\arxiv=1\centerline{\psfig{width=12cm,figure=biolattice.eps}}\fi
\qleg{Fig.\qhu 18b\qhv Formation of a network of euglenas
(seen from above).}
{The ``swimming pool'' of the euglenas was 12mm deep
and the upper layer had a thickness of 1.5mm (the role
of the depth with respect to the upper-layer remains
somewhat unclear). 
The authors do not give the level of light but as the 
photograph was strobe-illuminated the euglenas were not in
darkness. The authors do not say how long the network remained
unchanged.} 
{Source: Adapted from Plesset and Winet (1974).}
\end{figure}

Plesset and Winet say that the onset on instability is characterized
by a pattern of falling fingers which is maintained by ``return
upward swimming'' of the organisms. They explain the 
upward swimming by the fact that the micro-organisms under
consideration (namely {\it Tetrahymena pyriformis}) have 
negative geotaxis that is to say a tendency to go 
upward. Yet, why should organisms which have such an upward
swimming tendency let themselves fall to the bottom once
they have reached the surface. In other words what we have
here is the assumption of a dual behavior. When they
fall to the bottom, the organisms behave like inert particles,
yet when they swim upward they behave like living organisms.
Does the transition from one state to the other not require
an explanation?
\qpar

The explanations set forth by Harold Wager (1911) 
to account for the patterns he observed with {\it Euglena
viridis} raise the same difficulty. Just as {\it Euglena
gracilis}, {\it Euglena viridis} has positive phototaxis
(i.e. it is attracted by light) up to a given intensity threshold
above which it has negative phototaxis. Yet, this property played
no role in most of the experiments done by Wager because they
were performed in darkness. Light was applied only for taking
pictures. Consequently, phototaxis cannot be invoked
to account for the downward or upward movements of the euglenas. 
\qpar

On the contrary, in more recent expriments (e.g. Suematsu 2011)
the phototactic property of the euglenas plays a role because
a strong light is applied under the sample. Thus, the upward
movement is easily explained through negative phototaxis
whereas the downward movement may possibly be explained by
another effect such as self-shading or the fact that 
(for some reason) the euglenas
stop swimming
\qpar

The experiment reported at the beginning of this section 
according to which a lattice pattern appears (i) in darkness
and (ii) in a swimming pool of depth 0.1mm or 0.5mm suggests 
that one needs another explanation than bioconvection.
A depth of 0.1mm means that there are at most 2 or 3 layers
of euglenas. In other words such a case can probably
be approximated as a 2-dimensional problem. As in addition 
no exogenous factors (sich as light)  needs to be taken into
account, the situation is much simpler than bioconvection.
Moreover, as the lattice remains unchanged for several hours
it is legitimate to assume a stationary state.

\qA{Two compartment experiment}
A two-compartment experiment with ants was described earlier
in the section about clustering. It is natural to try it also
with euglenas. The figure below describes the device that was
used%
\qfoot{In fact we used the same device for the lattice experiment
because one experiment was in fact the continuation
of the other. For lattice formation
the fact of having two compartments played no role. It had
just the advantage of giving two
parallel observations instead of a single one.}%
. 

\begin{figure}[htb]
\ifnum\arxiv=0\centerline{\psfig{width=10cm,figure=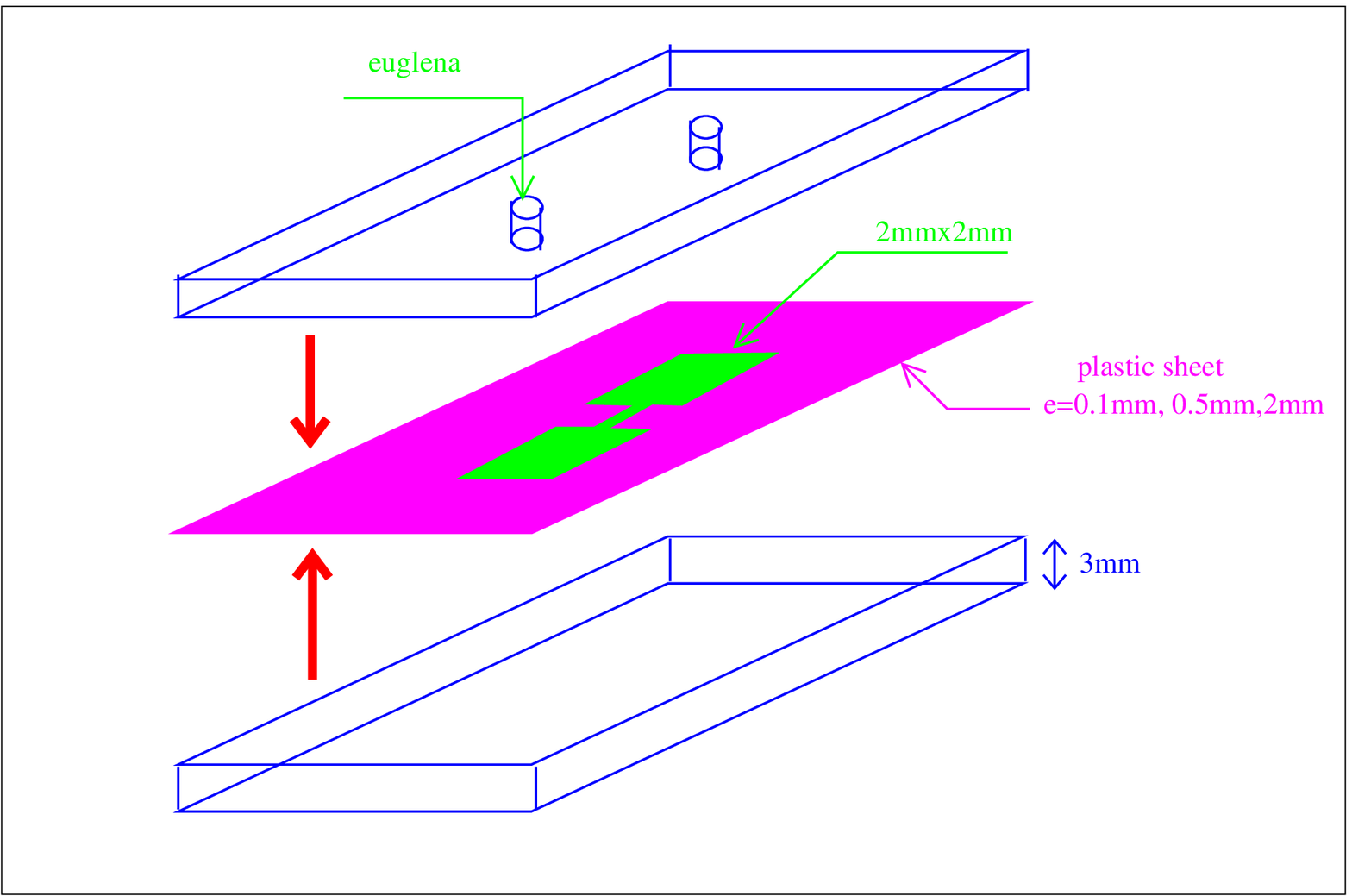}}\fi
\ifnum\arxiv=1\centerline{\psfig{width=10cm,figure=sandwich.eps}}\fi
\qleg{Fig.\qhu 19\qhv Size and sandwich structure of 
the two-compartment swimming pool.}
{} 
{}
\end{figure}

The main challenge was to fill the swimming pool 
with euglenas but without air bubbles. The technique depends on
the depth. Thus, for a depth of $ e= 0.5 $ mm the droplets
injected through one hole spread to the hole
pool, whereas for $ e=0.1 $ mm increased friction hinders 
the spreading. In this case the euglenas must be introduced before
the sandwich is closed; a layer of grease will prevent
any leakage in this first phase. Once the device was ready with
the euglenas inside the two holes were closed to prevent 
any evaporation. 
\qpar
The experiment was tried for 3 depths, namely $ e=0.1,0.5,2 $ mm.
It led to the following observations.
\qbu Immediately adter the device was filled and ready it was put
in complete darkness under a metal box. 
In order to make an observation the box was opened at time 
$ t=15 $mn and $ t=30 $mn. For $ e=0.1 $mm and $ e=0.5 $mm 
a lattice formed,
as already described. For $ e=2 $mm no lattice formed.
\qbu Once formed the lattice remained identical for several hours. 
\qbu In the following hours 
(in most experiments this was during night-time)
the lattice disappeared and was replaced
by a smooth and uniform distribution. Although the 
distribution was fairly uniform on each side, the concentration
was not the same. Most
of the euglenas were located one side.\qL 
On the contrary, for
$ e=2 $mm the density remained the same on each side.
\qpar

In the following weeks these experiments will be repeated
with special attention given to the role of the initial concentration
of the euglenas.

\vskip 5mm

{\bf Acknowledgements} \quad This work is basically a collective
undertaking. This is already clear from the fact that it has several
co-authors, but there were many other persons who provided help
and advice whose names do not appear in the co-author list.
We would particularly wish to thank
Prof. Hiraku Nishimori of Hiroshima University
who shared with us his enthusiasm for this
kind of research and Dr. C\'eline F\'erard 
of the IMPMC (University Pierre and Marie Curie) whose help
was invaluable.

\vfill\eject

\vskip 10mm
{\bf References}

\qparr
Costa (J.T.) 1997: Caterpillars as social insects.
American Scientist 85,150-159.\qL
[The paper has the following subtitle.
``Largely unrecognized, the gregarious behavior of caterpillars
is changing the way entomologists think about social
insects''. Needless to say, caterpillars are just one
example of insects displaying a gregarious behavior.]

\qparr
Dunoyer (L.) 1911: Sur la r\'ealisation d'un rayonnement
mat\'eriel d'origine purement thermique. Cin\'etique
exp\'erimentale [Realization of a molecular beam 
produced solely by thermal effect. Experimental kinetics].\qL
Le radium 7,142-146.\qL
[This article describes the first effusion effect
that is to say the propagation through 
vacuum of molecules whose velocities are nearly parallel
to one another.
In this experiment the mean free path of sodium molecules
at a temperature of 400 degree Celsius was about 20cm.]

\qparr
Economou (G.), Heidman (J.L.), Tsonopoulos (C.), Wilson (G.M.) 1997:
Mutual solubilities of hydrocarbons and water. III. 1-hexene,
1-octene, C10-C12 hydrocarbons.
American Institute of Chemical Engineers (AICHE) Journal 43,2,536-546.

\qparr
Einstein (A.) 1901: Folgerungen aus den Capillarit\"atsercheinungen
[Conclusions drawn from the observation of capillarity phenomena].
Annalen der Physik 4,3,513-523.\qL
[This was a paper in what became later known as chemical
physics.] 

\qparr
Engelmann (T.W.) 1879: \"Uber Reizung contractilen Protoplasmas durch
pl\"otzliche Beleuchtung. 
Archiv f\"ur die gesammte Physiologie des Menschen und der Thiere,
[also called Pfl\"uger's Archiv]
19,1-6 and 7-14.

\qparr
Engelmann (T.W.) 1882: \"Uber Licht- und Farben perception niederster
Organismen. 
Archiv f\"ur die gesammte Physiologie des Menschen und der Thiere,
[also called Pfl\"uger's Archiv]29,387-400.

\qparr
Gordon (D.) 2010: Ant encounters. Interaction networks and colony
behavior. Princeton University Press. Princeton.

\qparr
Holmberg (J.P.) Abbas(Z.), Ahlberg (E.), Hassell\"ov, Bergenholtz (J.)
2011: Nonlinear concentration dependence of the collective diffusion
coefficient of $ \hbox{TiO}_2 $ nanoparticle dispersions.
Journal of Physical Chemistry C 115,13609-13616.\qL
[The paper gives a clear example where the diffusion coefficient
depends upon the particle concentration.]

\qparr
Johnston (W.B.Jr.), King (J.G.) 1966: Measurement of velocity
distribution of atoms evaporating from liquid helium II.
Physical Review Letters 16,26,1191-1193.\qL
[Because the experiment was done near the boiling temperature
of helium it allowed a comparison between 
velocity distributions when the source was either
gas or liquid. There is a discrepancy between what is
predicted by the Maxwell-Boltzmann distribution and the experimental
results but its interpretation seems unclear.]  

\qparr
Lammert (B.) 1929: Herstellung von Moleculatstrahlen einheitlichen
Geschwindigkeit [Production of molecular beams of uniform
velocity].
Zeitschrift f\"ur Physik 56,244-253. \qL
[The author was a student of Professot Otto Stern in Hamburg and this
work constituted his thesis.]

\qparr
Lecomte (J.) 1949: L'inter-attraction chez l'abeille.
Comptes Rendus de l'Acad\'emie des Sciences (CRAS), 229,857-858.
[Attraction phenomena among bees. Paper published in the
Proceedings of the French Academy of Sciences.]

\qparr
Lecomte (J.) 1950: Sur le d\'eterminisme de la formation
de la grappe chez les abeilles [About the clustering process
in bees].
Zeitschrift f\"ur vergleichende Physiologie [Journal
for comparative physiology] 32, 499-506. 

\qparr
Lecomte (J.) 1956: Nouvelles recherches sur l'inter-attraction
chez {\it Apis mellifica}%
\qfoot{{\it Apis mellifica} is the same species as {\it Apis mellifera}.
In Latin the meaning
of ``mellifera'' is ``to bear honey'' whereas the meaning of 
``mellifica'' is ``to make honey''}%
.
[New results about attraction phenomena among {\it Apis mellifica}
(that is to say western honey bees).] 
Insectes Sociaux 3,1,195-198.

\qparr
Lucia (A.), Bonk (B.M.) 2012: Molecular geometry effects and
the Gibbs-Helmholtz constrained equation of state.
Computers and Chemical Engineering 37,1-14.

\qparr
Mast (S.O.) 1911: Light and the behavior of organisms.
John Wiley, New York.
[Several sections are devoted to {\it Euglena Gracilis}.]

\qparr
Nishimori (H.) 2012: Behavior of foraging ants under conflicting
information. Mini-workshop on collective behavior of social
insects and related topics. Hiroshima University, 6 July 2012.

\qparr
Pedley (T.J.), Kessler (J.O.) 1992: Hydrodynamic phenomena in
suspensions of swimming microorganisms. Annual Reviews of Fluid
Mechanics 24,313-358. \qL
[This is a long review paper which contains a few photographs
and about one hundred references.]

\qparr
Phillies (G.D.J.) 1974: Effects of intermacromolecular interactions on
diffusion. I. Two-component solutions.
The Journal of Chemical Physics 60,3,976-982

\qparr
Phillies (G.D.J.), Benedek (G.B.), Mazer (N.A.) 1976: Diffusion
in protein solutions at high concentrations: a study by
quasielastic light scattering spectroscopy. 
The Journal of Chemical Physics 65,5,1883-1892.

\qparr
Plesset (M.S.), Winet (H.) 1974: Bioconvection patterns in swimming
microorganisms cultures as an example of Rayleigh-Taylor
instability. Nature 248, 441-443.\qL
[This seems to be one of the earliest papers in the 
investigation of bioconvection based on the Rayleigh-Taylor
instability. It can be noted that such a 
mechanism was already suggested by
Wager (1911) although he did not explicitely mention Rayleigh-Taylor. \qL
In the present experiment, the depth was 12mm and the length
of the microorganisms was similar to the length of {\it Euglena
gracilis} that is to say about 50 micrometer.
Started in the 1970s, this stream of  research extended well
into the 1990s and 2000s.]

\qparr
Rabani (A.), Ariel (G.), Be'er (A.) 2013: Collective motion
of spherical bacteria. Plos One, 8,13,1-8.

\qparr
Roehner (B.) 2004: Coh\'esion sociale. Odile Jacob, Paris.

\qparr
Roehner (B.M.) 2005: A bridge between liquids and socio-economic
systems. The key-role of interaction strengths.
Physica A, 348,659-682.

\qparr
Schneider (W.R.Jr.), Doetsch (R.N.) 1977: Temperature effects on
bacterial movement. 
Applied and Environment Microbiology 34,6,695-700. 

\qparr
Stahl (E.) 1880: \"Uber den Einfluss von Richtung und St\"arke der
Beleuchtung auf einige Bewegungserscheinungen im
Pflanzenreiche. Botanische Zeitung (April): 298-413.

\qparr
Suematsu (N.J.), Awazu (A.), Izumi (S.), Noda (S.), Nakata (S.),
Nishimori (H.) 2011: Localized bioconvection of {\it Euglena}
caused by phototaxis in the lateral direction.
Journal of the Physical Society of Japan 80,064003,1-8.

\qparr
Van den Broeck (C.), Lostak (F.), Lekkerkerker (H.N.W.) 1981:
The effect of direct interactions on Brownian diffusion.
Journal of Chemical Physics 74,3,2006-2010.

\qparr
Viswanathan (G.M.), Luz (M.G.E. da), Raposo (E.P.), 
Stanley (H.E.) 2011: The physics of foraging. An introduction
to random searches and biological encounters. 
Cambridge University Press, Cambridge.

\qparr
Wager (H.) 1911: On the effect of gravity upon the movements and
aggregation of {\it Euglena viridis}, Ehrb., and other
micro-organisms.
Philosophical Transactions of the Royal Society, London,
Serie B 201, 333-390. \qL
[This is an early paper on bioconvection. As it is a long paper
(58 p.) which contains many pictures and drawings one expects
a fairly comprehensive description. For
{\it Euglena viridis} which is of the same size as {\it
Euglena gracilis} aggregation seems to occur only in
darkness which is the opposite of what is observed in
{\it Euglena gracilis}. This is perhaps related to the fact
that the threshold for the shift
from  positive to negative phototaxis seems to
be much higher.\qL
A great number of experiments
are described, but usually they do not have a well defined
purpose and their description is rather loose. Again and again
one reads that the experiment was done in a ``shallow vessel''
without any depth indication. \qL
The same remark
applies to the conclusion: ``The regularity of the aggregation
depends upon the depth of
the cell and the number of organisms present. It is more regular in a
shallow vessel than in a deep one''. That conclusion immediately
raises the question of whether there is a depth for which
the aggregation is maximum and whether
or not there is a critical
depth below which the aggregation disappears. Unfortunately, 
these are questions that the author did not address.] 

\qparr
Williams (C.R.) 2009: Pattern formation and hydrogen production
in suspensions of swimming green algae.
PhD Thesis, University of Glasgow.\qL
[This thesis is available on line; it is not about {\it Euglena gracilis}
but about two similar microorganisms, namely {\it Chlamydomonas 
nivalis} and {\it Chlamydomonas reinhardtii}. In contrast with 
the euglenas which are cylindrical, those organisms are rather
spherical. However, one would expect that the mathematical
methods developed for this case can also be applied to the
euglenas.]

\qparr
Yang (M.) 2011: Measurement of oil in produced water. in
Lee (K.) and Neff (J.) editors: Produced water.
Springer Verlag, Berlin.

\qparr
Zartman (I.F.) 1931: A direct measurement of molecular velocities.
Physical Review 37,383-391.

\end{document}